# Post-resonance reflections break the loss-vs-phase-resolution trade-off in microwave electronics

## Bala Govind[1] and Alyssa Apsel[1]

[1] Department of Electrical and Computer Engineering, Cornell University, Ithaca, USA

Corresponding author: bg373@cornell.edu

**Precise phase control of microwaves and millimeter waves is critical for today's wireless communication and signal processing electronics. For decades, achieving even modest phase-shifting resolution has demanded a complex mix of transistor switches and passive electromagnetic structures. These true-phase delay, true-time-delay or quasi-true-time-delay circuits operate near resonances that heavily attenuate and distort signals, limiting transmission bandwidth. Despite numerous attempts, passive phase shifters have remained lossy and incompatible with high-channel-capacity beamforming, irrespective of the semiconductor fabrication process. In this article, we introduce a mechanism whereby waveguide reflectors based on coupled resonances can be reprogrammed to evade loss. Central to their operation is that loss from internal resonances is confined to low frequencies, while at high frequencies, their reflectivity is maximized and broadband phase variations, induced by those resonances, still persist. Since performance in the low-loss post-resonance spectrum is largely agnostic to the number of switches, ultra-fine digital tuning is possible. This breaks the historical tradeoff between loss and precision to achieve resolution surpassing state-of-the-art integrated circuits by over three orders (bits) of magnitude. The device does not distort the transmitted signal and consumes no power. Moreover, the chip occupies a sub-wavelength footprint in a Complementary Metal Oxide Semiconductor platform. This makes it an optimal candidate for seamless integration in on-chip multi-gigabit data links, radio astronomy transceivers and control hardware for millimeter-wave qubits.**

Phase shifting is a quintessential capability of all high data-rate microwave, optical or acoustic signal processing hardware and wireless communication systems[1-10]. Arrays of phase modulators add signals coherently in beams directed to or received from a target. This scheme makes transmission resistant to attenuation and interference over extended distances in lossy media. The endeavor to integrate miniature transmitter and receiver arrays into cell phones and base-stations boosted the development of compact phase-shifting integrated circuits[11-15] which are integral to Fifth and Sixth Generation (5G and 6G) technologies[11-15] that reach frequency bands higher than 20 GHz. To scale well, these systems must adhere to the space constraints of a semiconductor chip, while minimizing power consumption, noise, distortion and, most importantly, loss of the signal. Through tight packing density and high phase resolution, these arrays should form beam patterns that do not spread into the environment. That is, a high Effective Isotropic Radiation Efficiency[16]. Unfortunately, thus far, these elements have been subject to what was assumed to be a fundamental tradeoff between phase resolution and the loss they impose. Fig 1a illustrates the problem. Conventional phase-shifters first attenuate and distort the transmitted information. This requires subsequent boosting of the signal via amplifiers which degrade linearity of the signal[17-19]. Therefore, to reduce loss from resonances and other switch-based tuning mechanisms and lessen the deleterious effects of amplifiers, most systems simply sacrifice phase resolution[20-24]. This compromise involving the use of coarse phase shifts, in turn, produces higher quantization noise [25,26] which would be manifested by stronger side lobes in the radiated electric field. This again degrades the transmission, necessitating sophisticated digital signal processing hardware for error correction. Consequently, most programmable phase shifters have a mediocre resolution (minimum phase shift exceeding 10°).

As Fig. 1b shows, through a technique we call "post-resonance tuning", we break the loss vs resolution trade-off by producing an ultra-fine phase resolution of under 0.2°, surpassing previously reported resolutions in all integrated circuit technologies while incurring only 5 dB of loss across an ultra-broadband spectrum. This is possible because the new method operates in a spectrum that lies beyond the resonances where remnants of their lossy effects due to microwave switches and electromagnetic structures have subsided, and where large phase shifts still linger. It nearly eliminates the need for amplifiers to compensate for attenuation and does not inject external nonlinearity into the original microwave transmission. A key advantage of the final device is that it consumes no DC power and occupies a sub-wavelength footprint (Fig 1c) in a Complementary Metal Oxide Semiconductor (CMOS) process, making it readily usable in consumer electronics. It, paradoxically, converts the low quality-factor of distributed transmission lines and capacitor banks in CMOS into an advantage by ensuring mild variations in phase and low-loss gradients across over 10 GHz of bandwidth.

## The loss-vs-phase-resolution problem

Before examining the mechanism of the post-resonance tuning scheme in the following section, it becomes necessary to grasp the implications of the stubborn loss-vs-resolution trade-off in microwave electronics. This is of importance in the design of cell phones and base-station transceivers that must integrate phase-shifting hardware for multi-Gigabit transmission. Eliminating the class of power-hungry active phase-shifters[17,18,19] leaves a suite of passive techniques to choose from. An extensive comparison of delay elements fabricated in CMOS, Gallium Nitride, Gallium Arsenide and Micro Electromechanical Systems is shown in Supplementary Table 1. Representative techniques for different phase-resolutions, and the post-resonance tuning scheme, are compared in Fig. 2a. We see the heretofore accepted compromise between phase-resolution and the loss they impose. Thus far, it was assumed that these elements could not be small, have low loss, and have high resolution at the same time. Each state-of-the-art device sacrificed one or more of these features for another. This trade-off stems from the low quality-factor of transistor switches and inductors offered by commercial chip foundries at millimeter-wave frequencies. The transistors' parasitic capacitances, in combination with on-chip inductors, produce lossy resonances that absorb the microwave signal and limit the usable frequency bands for communication. This becomes pronounced if longer delay lines are used in a true or quasi-true time delay[10] scheme with tunable return paths. They eventually encounter these resonances that limit their bandwidth and do not operate in a post resonance regime. This only worsens as more switches are added to produce finer phase resolution. This begs for a scheme that is agnostic to the number of switches added or, in other words, an insensitivity to the complexity of digital tuning needed for beam steering.

Since post-resonance tuning relegates the lossy resonances to lower frequencies, what lies higher than those is a broad, near-lossless band in which ultra-high-resolution phase-shift of under 0.2° is possible. Thereby, as Fig 2a also shows, this scheme cleanly breaks the trade-off, and that loss is indeed indifferent to the number of switches used in its circuit.

One metric that clarifies the strength of various delay circuits for microwave communication is the Array Factor. It is the sum of individual phase shifters' contributions when they are arranged in an array (see Methods Section 4 for a derivation). Array Factors for beams produced for three popular state-of-the-art techniques, and the post-resonance tuned scheme, are shown in Fig 2b. First, all-pass networks[23] could be considered. They are a class of broadband delay circuits in which one of a few options of fixed phase shifts is selected. They have low loss owing to few switches but are inherently coarse resolution devices. While, in principle, they may have small form-factors that result in good beam directivity, their coarse phase shift manifests itself as large sidelobes that heavily distort the radiated signal. Instead, to increase the bits of resolution, switched-filters[20] could be used. These, however, are composed of large capacitors and lossy inductors, and only a couple of phase-shifters can be accommodated on chip. Additionally, they are optimized for performance in the narrow frequency band around a resonance. These factors result in poor beam directivity in a target direction. Lastly, to increase the packing density in an array and to increase phase resolution to seven bits of tuning (4° of phase resolution), a popular Passive Vector Modulation[34] scheme could be tried. It comprises an active phase-shifter minus the gain element. Unfortunately, its insertion loss of over 15 dB is unacceptable for most practical applications. In contrast, the post-resonant tuning scheme not only accommodates sixteen elements within 1 $mm^2$, but also gives a near-infinite resolution of ten bits. The advantage is manifested, as Fig 2b also shows, for beams steered at a range of angles, wherein quantization noise and unwanted sidelobes is significantly suppressed. This produces the high signal-to-noise ratio needed to detect microwatt-level microwave signals in radio astronomy[38]. Notably, it achieves this with negligible insertion loss, which also makes it the best candidate for precise phase-coupling between millimeter-wave antenna arrays[35].

## The post-resonance reflection mechanism

We will now examine how the ultra-sharp-resolution post-resonance tuning also minimizes loss at microwave frequencies. To convert the one-port post-reflection tuned reflector into a two-port device i.e., a phase-shifter, we deploy it in a reflective scheme that utilizes two reflectors and a coupler and has gained traction recently in both microwave[22,27] and superconducting circuit designs[48]. This is shown in Fig 3a. The microwave signal is injected into the Input port of the quadrature hybrid coupler that splits the signal into two components with a 90° phase shift between them. These half-power components reflect off tunable reflectors at the Through (T) and Coupled (C) ports of the coupler. In this design, the coupler has a miniature size since it is implemented on two metal layers as overlapping coils of a transformer (Fig 3a. ii). This is in opposition to coupled wavelength-scale transmission lines[28-29].

Now, a decision must be made among infinite possible structures of the reflector. One that is maximally reflective, and which induces large phase shift is desirable. Any prototypical resonance behavior is characterized by phase shift. This has heretofore been the concept behind the operation of devices using resonances in reflective phase shifters[30-31] but does not consider the insertion loss and bandwidth of the device. That is, so far, delay was induced during the reciprocal exchange of energy between electric and magnetic fields in the resonant filter inserted on the coupler's ports. However, this mechanism induces a swift phase shift only within a narrow frequency band. Beyond this band, a substantial portion of the signal is absorbed by the reflector, leaving little signal to be reflected. We, instead of optimizing for a narrow resonant band, seek to operate in the wide swath of spectrum beyond the resonances that have limited the usable bandwidth of previously reported phase-shifters, quasi-true and true-time-delay elements (Supplementary Table 1).

In this system, as Fig 3b and Supplementary Fig 1 show, each reflector consists of three waveguide segments, with the final segment connected to ground to complete a return path for the microwave signal. Each waveguide segment is primarily inductive in nature. In addition, tunable capacitors are inserted intermittently at the junctions of these segments. This closely resembles a "traveling wave" structure formed of a non-uniform waveguide whose impedance can be changed internally by varying these capacitors. As Fig 1c shows, waveguide segments are synthesized from a distributed, coiled structure on the highest metal layer of the process stack, with tunable capacitors connected on lower metal layers. These digitally tuned capacitors are implemented as five-bit capacitor banks. Each capacitor bank, in turn, comprises several parallel legs of N-type Metal Oxide Semiconductor switches that bring metal-oxide-metal capacitors in or out of action (Extended Data Fig. 1d). The input and output ports of the full device are accessed through Ground-Signal-Ground-Signal-Ground (GSGSG) waveguides.

The $\pi$-shaped sections of waveguide segments and capacitors couple with one another to form multiple hybrid resonances. These induce programmable phase delays corresponding to the values of the tunable capacitances. Smaller capacitances lead to fewer pronounced resonances and give smaller phase shifts (Fig. 3b.i). Larger capacitance states produce multiple strong, coupled resonances that impart large phase shifts (Fig. 3b. ii). Supplementary Note 1 gives a direct formulation of reflection coefficients' magnitude and phase shift. Importantly, since the three-segment nonuniform transmission line is a waveguide, there exists an upper frequency limit ($f_{cutoff}$) beyond which no signal can propagate down its length, akin to the Bragg periodic cutoff frequency[36]. This broadband structure best suits a reflective-type delay element since it would be maximally reflective beyond this frequency.

As Fig 3c shows, the tunable capacitor-loaded waveguide segments produce numerous lossy, coupled resonances below 20 GHz. It is within this frequency range that we encounter fluctuations in the fraction of signal reflected from these waveguides. Also, these resonances generate broad dips in the amplitude response of the reflection coefficient (Fig 3c.i). This is because of the poor quality-factors of the circuit components that created them, as could be expected from CMOS process technologies. However, unlike loaded line phase shifters[46] for which optimal impedance matching between waveguide segments is required for transmission, the waveguides here rely heavily on impedance mismatch to behave as good reflectors. Here, we opt to explore the often-overlooked regime of the ultra-broadband spectrum above the inter-segment resonance frequencies. See Supplementary Note 2 for a derivation on how the periodic structure comprising the reflector can be modeled as a filter with finite Bloch impedance, and frequency-dependent propagation constants. Paradoxically, as variations in these metrics show (in Supplementary Figs 3 and 4), lower quality factors of under 5 are more conducive to a post-resonance operation since they facilitate slow variations in input impedance and retain a large tuning range of phase shift beyond the last coupled resonance. This is also supported through a derivation of scattering parameters of a chain of these resonators from generalized Coupled Mode Theory in Supplementary Note 3 and Supplementary Figure 5.

Importantly, the phase response of the reflection coefficient (Fig 3c. ii) reveals the advantage of repurposing multiple lossy low-frequency resonances. Tuning the natural frequencies ($f_1, f_2 \ldots f_n$) of the cascade of resonators within the reflector modulates the positions of the coupled resonances ($f_{coupled,1}$, $f_{coupled,2}$, … $f_{coupled,n}$), alike supermodes, thereby determining where the recovery in reflection coefficient begins. While the magnitude of the reflection coefficient simply experiences dips and recoveries within finite bandwidths around each resonance, a phase shift is accumulated after each coupled resonance. Consequently, beyond the last resonance at 20 GHz, the reflecting waveguides transmit no signal but only impart cumulative, broadband phase shifts. Essentially, isolating lossy resonances at lower frequencies enables the utility of as many switches as the bits of resolution needed for phase tuning (Fig 3d). This is possible because although each switch contributes to a signal-absorbing resonance, this local behavior is immaterial since the device wouldn't be operated at these low frequencies but rather in the broad spectrum higher than them. Conveniently, this makes optimal resonant filter design of segments $L_1$, $L_2$ and $L_3$ and capacitors $C_1$, $C_2$ and $C_3$ unnecessary.

As we indicated, finite quality factors of the switches manifest as internal dissipation rates, aiding in coupling resonators of different natural frequencies and facilitating a gradual transition in the phase curves between coupled resonances. This unintuitive preference of a low quality-factor over a high quality-factor in this technique is also supported by transitions in input (Bloch) impedance and phase shifts per section (propagation constant) of the composite waveguides as shown in in Supplementary Figs 3 and 4. It ensures distinct roll-offs beyond the highest coupled resonance frequency for each digitally tuned phase state. Specifically, when activated or deactivated, the switches alter the decay rates of the coupled modes in the multi-resonant waveguide, thereby tuning the ultra-broadband phase shift. At extremely high frequencies (beyond 200 GHz), the phase curves would gradually converge asymptotically, causing the reflector to behave akin to a short circuit to ground. Nonetheless, for all practical purposes, distinct phase shift is achieved across tens of gigahertz within the band for all digitally tuned states.

In the physical design of this delay element, to limit the total tuning bits to ten, capacitor banks $C_2$ and $C_3$ were digitally tuned to identical values (i.e., $C_2 = C_3$) for every phase-state. Further, the discrete inductor $l_{\text{ext}}$ (Fig 1c) is simply an appendage of the waveguide segment $l_3$ that helps boost the phase tuning range beyond 180°. Therefore, tuning the five-bit capacitor bank $C_1$ (see Extended Data Fig 1d) inserted close to the beginning of the waveguide and the two other five-bit capacitor banks downstream (equal to $C_2$) determines the frequency at which the last resonance occurs and, thereby, where the recovery in insertion loss begins. This approach represents a departure from the capabilities of conventional microwave electronics since, normally, using ten or more switches in a tunable impedance would prohibitively distort and weaken a signal.

## Characterization of the post-resonance reflection-based delay

The concept of the post-resonance tuning scheme is validated through two-port measurements of a CMOS chip that was implemented on a Fully Depleted Silicon-on-Insulator 28 nm platform, within an area of only 0.064 $mm^2$ in Silicon. See Methods sections 1 and 2 and Extended Data Fig 3 for a description of the design methodology and experimental setup. In this technology, the leakage of electromagnetic field from the passive structures to the substrate is negligible, sustaining performance at millimeter-wave frequencies. The chip was designed to operate in the 5G New Radio spectrum for Multiple-Input-Multiple-Output (MIMO) communications. The broad frequency range of the circuit is attributed to the broad bandwidth of the coupler and distributed transmission line segments.

Fig 4a shows the spread of both insertion loss and phase tuning for over 800 measured states of the phase shifter. We see that loss recovers beyond the sub-20 GHz band, for all states. This is because the frequency of 20 GHz is close to the cutoff frequency of the reflective waveguides, beyond which no transmission can occur, and the signal is reflected back. In this regime, the loads on the Through and Coupled ports exhibit maximal reflectivity.

Narrowing our attention to 20-30 GHz frequency range, we see that this scheme produces over 200° of phase tuning (Fig. 4b.1) with an ultra-fine phase resolution of under 0.3° (Fig 4b.2) across the band. In practice, if more than 360° is required, a simple phase-flip circuit could be included [32]. In Methods section 2 and Extended Data Figs 1a, b and c, we show how this range and resolution can be further improved by altering the length of the final transmission line segment and by adding a true-time-delay. The extension in length may be made within the original footprint, as the photograph of a variant of the chip in Extended Data Fig 1e shows. Alternatively, if 9 to 10 dB of loss (by previously reported phase-shifters) is tolerable, cascading two post-resonance tuned phase shifters could yield over 400° of phase tuning range.

Crucially, the loss through this phase shifter reduces with increasing frequency within the broad 21-30 GHz frequency band (Fig 4c) and, on average, is about 4.9 dB across the band. This contrasts with the conventional trend in high-frequency electronics and transmission line measurements wherein loss worsens at high frequency due to poor quality factors of switches and passive elements. This confirms that the magnitudes of reflection coefficients of these loaded transmission lines progressively increase with frequency, as we postulated earlier. Fig. 4d shows that the return loss (the percentage of signal reflected to the ports) exceeds 15 dB for every state of the phase shifter. This indicates that altering the reflective load has negligible effects on the coupler's coupling coefficients to its ports.

Since no analog-voltage-tuned varactor diodes[27, 33] were used, the component's ability to reproduce the signal without harmonic distortion is boosted, far more than current microwave techniques. To investigate this, the power of intermodulation products was measured across frequency and for high and low delay states. The device was found to have a power at which intermodulation products matched the power of the fundamental tone (an input referred $IP_3$) of over 25 dBm. This corresponds to a power at which the gain compresses by 1 dB (input referred $P_{1dB}$) of over 15 dBm, higher than all previously reported CMOS phase shifters' signal linearity metrics[24]. This is further attributed to the mode of operation not being near dispersive/distorting resonances and to the Fully depleted Silicon-on-Insulator switches conducting very linearly. See Extended Data Fig 4 for these measurements.

## Outlook

In this article, we have introduced a method that breaks the traditional tradeoff between loss and angular resolution in phase-shifting elements. Unlike power-inefficient active phase shifters which have become the workhorse of commercial integrated phased arrays, we offer a device that operates without consuming any DC power and provides almost distortion-free transmission. The core idea uses phase shifts from resonances in an unconventional manner. By placing unwanted, lossy resonances at lower frequencies, reflectors are made virtually lossless at higher frequencies, while retaining ultra-wideband phase shift. Their 9 GHz of bandwidth and record-high resolution (0.2°) makes unnecessary the convention of combining multiple phased arrays that each service smaller frequency bands. Vitally, post-resonance tuned reflectors do not use analog voltage-tuned varactor diodes that distort the information transmitted or received and, thereby, have higher signal linearity than previously reported circuits. This would potentially eliminate the complicated calibration hardware and back-end digital signal processing previously deemed essential for weakly linear delay elements as well.

We have shown that sub-wavelength non-uniform transmission lines used as reflectors could facilitate up to 16 elements in just 1 $mm^2$ of chip area. This would boost the directivity of beams, for stronger reception. Their 10-bit resolution, supplemented by amplitude weighting, would rapidly suppress sidelobes away from the main lobes to eliminate production of phantom-like artifacts that blur images in high bandwidth systems like Frequency Modulated Continuous Wave Radar[37]. Significantly, the scheme is implemented in a commercial CMOS platform, which quickly opens up possibilities of beamforming for peer-to-peer communication and edge-computing with handheld devices. Its ultra-compact footprint could also enable the integration of unconventional systems like interferometry-based ambient health sensors[39] and Reconfigurable Intelligent Surfaces[40] in the small form-factors of cell phones or tablet computers. Given their compact size and resolution, post-resonance reflection-based delay elements could also find application in accurate phase control of drives for qubits[41,47] at millimeter-wave frequencies.

## Data availability.

All measured data that supports the conclusions in this investigation are available in the main article, supplementary information and at Zenodo (https://zenodo.org/records/10723910).

## Code availability.

Code for comparing projections of Array Factors from different phase-shifting mechanisms and basic coupled-mode formulation is at the code repository (https://zenodo.org/records/10723910).

## Methods

### 1. Chip design and electromagnetic simulation.

The chip was fabricated by ST Microelectronics (France). Design was done using the N-Type Metal Oxide Semiconductor transistor models provided by the foundry for its 28nm FDSOI CMOS process. Circuit level simulations were performed using Siemens's Calibre tool within Cadence™'s Virtuoso environment. Electromagnetic simulation of the phase shifter comprising tunable capacitor-loaded transmission line segments, the hybrid coupler and Ground-Signal-Ground-Signal-Ground waveguides was performed with the Momentum 2.5D tool, offered by Keysight™. Small signal transfer parameters of this structure were used in co-design with the transistor switches' device models, from which parasitic resistances and capacitances were extracted. Harmonic Balance analysis[42,43] was used for large signal linearity analysis.

### 2. Extending the phase tuning range and resolution of the primitive structure of the reflective loads

The scheme of using post-resonance reflections was used in the main article to design a phase shifter that had minimal loss and over 180° of phase tuning. In some applications, an even larger phase tuning range may be required. In other instances, a higher resolution could be useful. Here, we show how to extend the tuning range and available bits, without consuming extra chip area. Consider the diagram of the modified reflector in Extended Data Fig. 1a. It differs from the primary structure in Fig. 3 by only the extension in length of the return path of the microwave signal. In addition to the segments $l_1, l_2$ and $l_3$, there is an option to bring an additional segment $l_4$ into action. It is activated by switches $S_1$ and $S_2$. When these switches are turned off, a fourth $\pi$-shaped resonator, formed of the inductive segment $l_4$ and parasitic capacitances of these switches, cascades with the original structure, producing a new set of coupled modes. These states that have moderate phase shift. When the switches are turned on, however, a true-time-delay through $l_4$ is added to the original phase states of the primary three-segment structure (Extended Data Fig. 1b) to give large phase shifts. The circuit model that represents the combination of coupled resonances of the original structure and the true-time-delay of the final section is shown in Extended Data Fig. 1c. For clarity, the structure of the binary-weighted capacitor banks that facilitates fine tuning of these coupled resonances, is shown in Extended Data Fig. 1d.

To demonstrate the concept of combined phase and time delay in the same structure, a variant of the original CMOS chip was designed. Its photograph is shown in Extended Data Fig. 1e. Here, to accommodate the longer return path in the modified reflector, the waveguide segment $l_4$ is fabricated as an extension to the original coil comprising segments $l_1, l_2$ and $l_3$. It is inserted within this coil as a new turn. This boosts the mutual inductance between the coils. Switches $S_1$ and $S_2$ are large mm-wave transistor switches that can be activated to short the signal to ground.

The measured spread of achievable phase-shifts and corresponding insertion loss with respect to frequency is shown Extended Data Fig. 2a. The phase tuning range is increased by over 30°, to 231° as Extended Data Fig. 2b.i shows. This variant of the basic post-resonance tuned phase shifter increases the number of bits to 11 (< 0.17°) of fine-tuning phase resolution (Extended Data Fig. 2b. ii), the highest reported resolution for digitally tuned phase shifters. As in the original structure in the main article, reflection coefficient of the loads recovers after the cutoff frequency of the nonuniform transmission line, $f_{cutoff}$, and the insertion loss is minimized post resonance to an average of 5.5 dB (Extended Data Fig. 2c). Finally, as Extended Data Fig. 2d shows, the return loss is over 20 dB. This implies near-ideal broadband matching to the input and output ports for this variant of the chip.

### 3. Measurements

The experimental setup to measure input-output phase delay characteristics and the signal linearity is shown in Extended Data Fig. 3. The CMOS chip was wire-bonded to a Printed Circuit Board. DC voltage bias and digital bits were fed to the board from a Digilent Analog Discovery static I/O module. Small-signal RF performance was characterized on a microwave probe-station, through de-embedded S-parameter measurements with an Agilent 8722ES 40 GHz Vector Network Analyzer. To measure signal linearity, two Anritsu 68369B RF synthesizers produced fundamental tones and the third-order intermodulation products (IIP3) were measured using an Agilent 8564EC spectrum analyzer. Since the RF power output beyond 20 GHz was limited to about 12 dBm in both signal sources, 1 dB gain compression points (P1dB) were not directly measured. Instead, input- referred gain compression was calculated from its theoretical relation to IIP3 measurements[44]. These are shown for cases of minimum and maximum phase shift and across frequency in Extended Data Fig. 4.

### 4. Array Factors for phase shifters with finite phase resolution

The Array Factor (AF) for any microwave, optical, or acoustic transducer array plays a crucial role in determining the accuracy of beamforming, directing radiation in a programmed direction[45]. We have previously considered, more generally, the calculation of this metric in delay elements that produce minimal beam squint across wide bandwidths in the Methods section of Ref. 10. However, in a phased array, since the elements produce only pure phase shift rather than a broadband time delay, the calculation of the sum of individual elements' contribution is straightforward. When arranged in one-dimensional linear array, assuming that the inter-element separation, $d_x$, equals λ/2, where λ is the midband frequency of the array and that the array has infinite phase resolution and no insertion loss,

$$AF_{\text{phased array, ideal}}(\theta) = \sum_{n=0}^{N-1} a_n \exp\left(j\omega n\left[\delta_{(\theta)} - \frac{\omega_0}{\omega}\tau_0\right]\right) \tag{1}$$

$a_n$ is the weight assigned to the n[th] element and for simplicity, identically equals the insertion loss (IL) in forthcoming analysis. Here, $\omega_0$ is the center-frequency, $\omega$ is the frequency of radiation in the incoming beam. $\tau_0$ is the effective time increment to align the beam to a broadside angle of $\theta$. $\delta_{(\theta)}$ is the angle-dependent "projected delay". It equals $\frac{d_x}{c}\sin\theta$, with $c$ being the speed of light. However, as the main article presents, all delay elements have a finite resolution and insertion loss. Since phase is quantized by $B$ bits, there are $2^B$ possible phase states. Then, the quantized phase that each element must be programmed to is

$$\alpha(n, \theta, B) = \frac{2\pi(n-1).\,d.\sin\theta}{2^B} \tag{2}$$

The adjustment to eqn. 1, including insertion loss (*IL*) is

$$AF_{\text{phased array, actual}}(\theta) = \sum_{n=0}^{N-1} IL_{\text{phase-shifter}} \exp\big(j \cdot \alpha(n, \theta, B)\big) \tag{3}$$

This formulation is used for the comparison shown between arrays of passive phase-shifters in Fig 2b of the main article. For simplification, these quantized phase shifts are rounded in the "quantize_phase.m" function in the code depository.

## Acknowledgements

The authors acknowledge the Cornell NanoScale Facility, a member of the National Nanotechnology Coordinated Infrastructure (NNCI), which is supported by the National Science Foundation (Grant NNCI-2025233) and where the work was done in part. B.G. thanks T. Tapen for on-chip microwave calibration and F. Wu, F. Monticone and S. Huang for discussions in preparing the manuscript. Both authors thank A. Cathelin from ST Microelectronics for chip fabrication.

## Author contributions

B.G. developed the concept of broadband post-resonance tuning and designed and tested the CMOS chip. A.A supervised the experiments and development of theory. B.G. wrote the manuscript, with input from A.A.

## Competing interests

The authors have filed a US provisional patent based (Application number: 63/633,380) on the post-resonance reflectors.

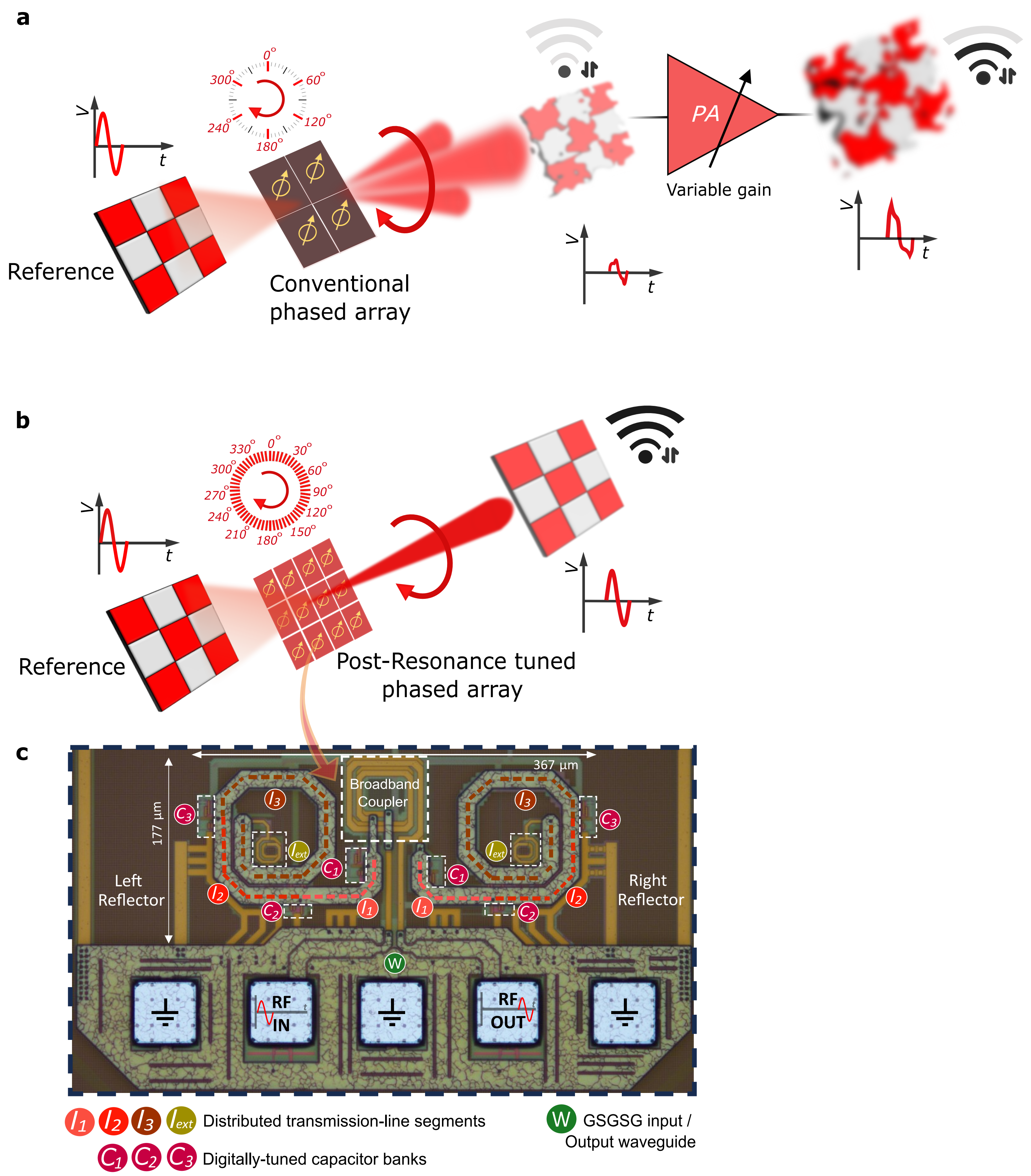

**Figure 1. Maximum phase resolution with minimal microwave transmission loss achieved through post-resonance reflection.** **a,** Traditional integrated-circuit phase-shifters rely on resonance-induced delay. They are bulky and have coarse phase resolution. Arrays of these elements yield broad, distorted beams. To counteract loss, variable gain power amplifiers (PAs) are used, but they introduce additional distortion. **b,** In contrast, post-resonance tuned phase shifters offer a compact, sub-wavelength footprint, generating pencil-thin beams from dense arrays. With 10-bit phase resolution, they effectively eliminate sidelobes and boost signal-to-noise-ratio. Their low loss reduces the need for amplification, preserving the integrity of information transmitted in Gigabit wireless links. **c,** Each element within the proposed array features input and output ports accessed through a Ground-Signal-Ground-Signal-Ground (GSGSG) waveguide. Fabricated in a 28 nm CMOS platform, this device utilizes a miniature coupler to divide the microwave signal, directing it towards identical reflectors. Each reflector is loaded non-uniformly with digitally-programmable capacitor banks, creating coupled resonances at low frequencies and finely-tuned, broadband phase shifts at higher frequencies. The waveguides' sub-segments are constructed from coiled traces on three metal layers, resulting in an ultra-compact form factor.

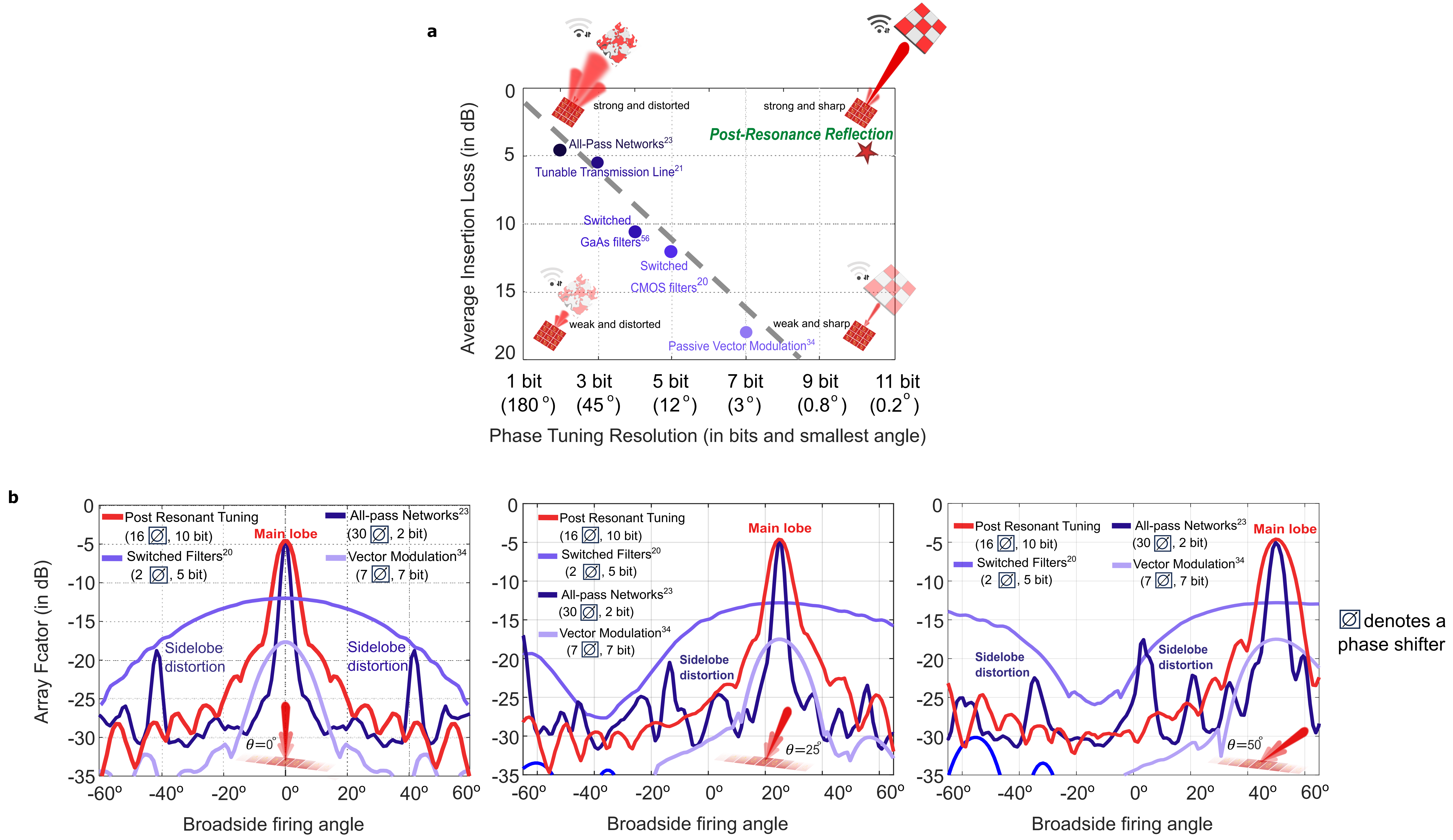

**Figure 2. Post-resonance tuned phase shifters break the tradeoff between loss and resolution. a,** In conventional microwave delay elements, loss escalates linearly with the number of tuning bits. This is because of poor quality factors of transistor switches, varactors, inductors and waveguides. State-of-the-art delay element circuits face strong resonances, inherently limiting their bandwidth. To address this issue, they utilize only a few phase-delay paths, sacrificing phase resolution. In contrast, post-resonance tuning achieves minimal loss of 5 dB while offering the sharpest phase resolution of 0.3°. **b,** Microwave beam directivity compared in terms of Array Factors (gain) for 0°, 25° and 50° broadside firing angles. Passive methods encounter challenges: (i) smooth Array Factor but low directivity due to poor packing density (Switched filters), (ii) good directivity but distorted beam profiles due to poor phase resolution (All-Pass Networks), or (iii) high resolution but significant signal attenuation (Passive Vector Modulation). Post-resonance tuned phase-shifters have sub-wavelength footprints for high packing density and exhibit minimal loss. They significantly suppress sidelobes because of their near-infinite resolution.

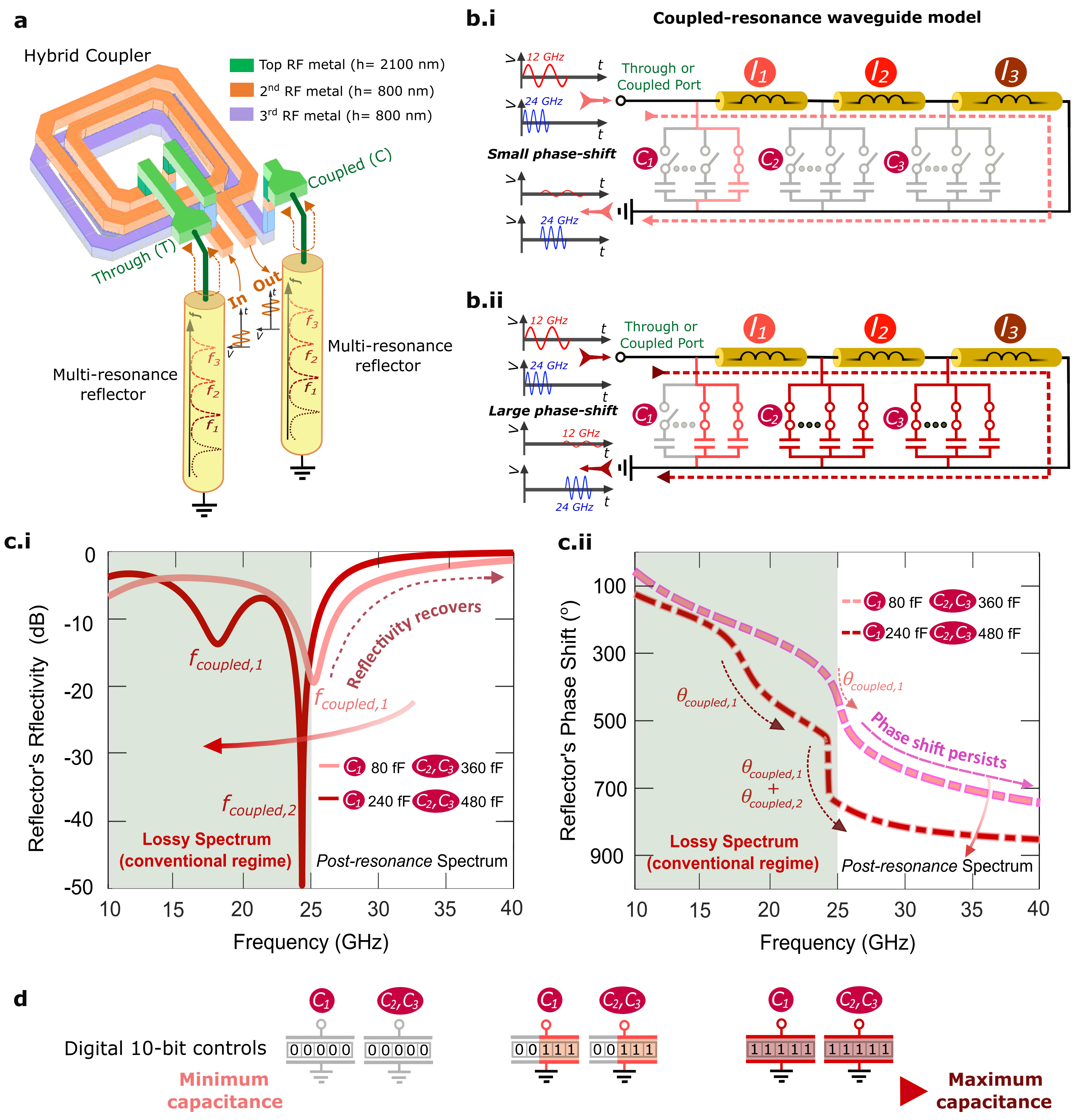

**Figure 3. Post-resonance tuning extends beyond the lossy resonant spectrum to achieve broadband phase tuning. a,** Mechanism of the reflective type phase shift. Signals bounce off of waveguide loads on a broadband coupler's Through and Coupled ports. They add at the output port with a common phase shift. **b,** One reflector, of a pair, generates multiple coupled resonances using inductor-like sub-segments and digitally-tuned capacitors.**(b.i)** A small phase-shift is achieved by activating fewer capacitors with switches. **(b.ii)** A large phase-shift is achieved by activating more capacitors in these banks. **c,** One or more coupled resonances ($f_{coupled,1}$, $f_{coupled,2}$ ...$f_{coupled,n}$) arise from combinations of waveguide-segments and capacitor bank settings. **(c.i)** The frequency response of the reflectivity (reflection coefficient) shows large signal absorption at low frequencies, with recovery occurring beyond the final resonance. Larger capacitances yield multiple prominent coupled resonances. (**c.ii**), Conventional methods use phase shifts induced by resonances in the lossy low-frequency regime. Instead, utilizing the enduring phase shift post-resonance is beneficial, as loads are maximally reflective. In this illustration, waveguide sub-segments $l_1$, $l_2$ and $l_3$ have identical inductances of 300 pH and $C_3$ equals $C_2$. **d,** The digital tuning scheme wherein control bits activate smaller capacitors within banks $C_1$ , $C_2$ and $C_3$ to aggregate capacitance.

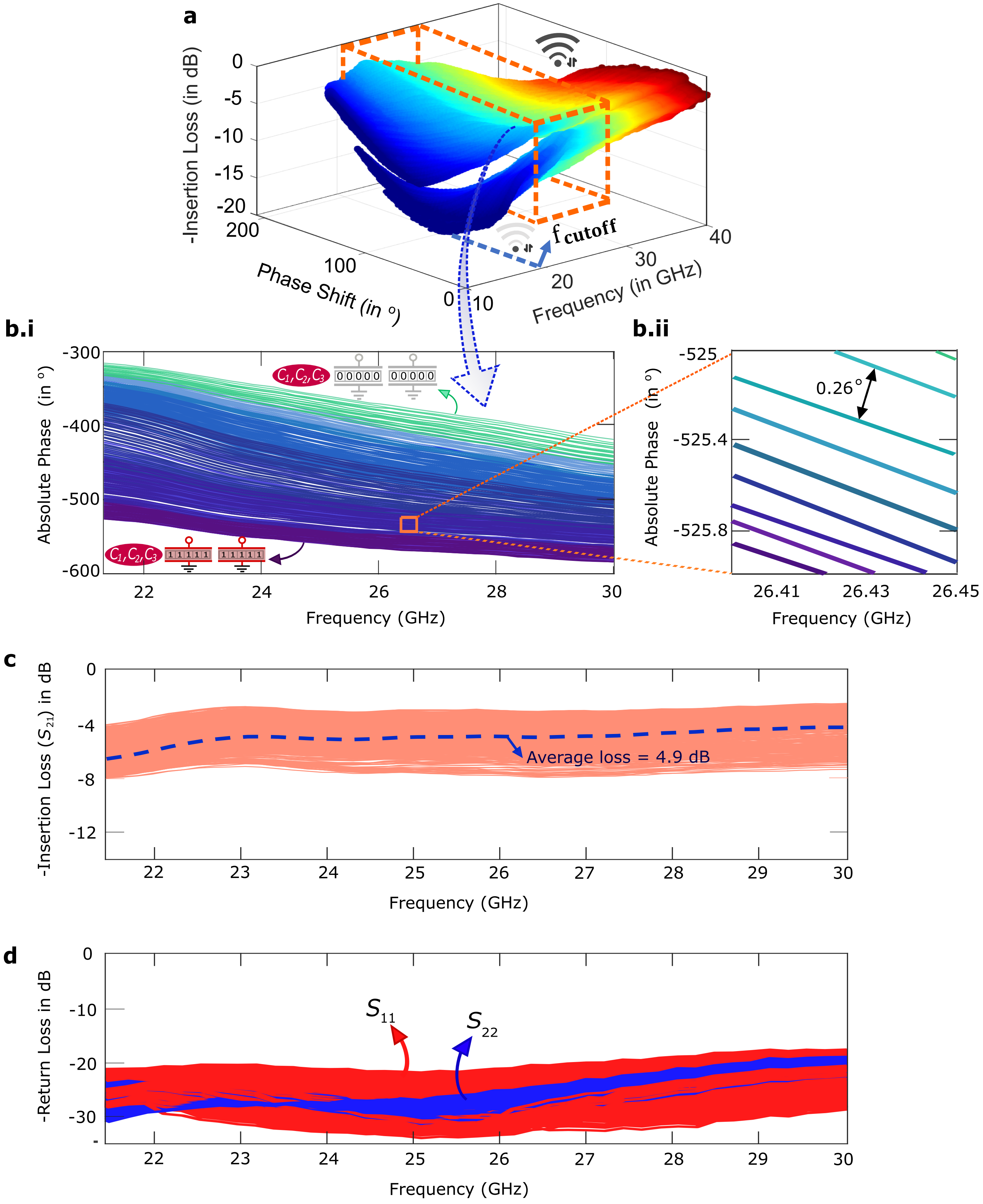

**Figure 4. Measured phase shifts, fine-tuning resolution, insertion loss and return loss for 800 states of the post-resonance tuned delay element. a,** Spread of phase-shift and loss for 10-bit controls of the digitally-tuned capacitor banks integrated in the non-uniform waveguides. Loss reduces at higher frequencies, as predicted by Coupled Mode theory (see Extended Data Fig 5). Lossy, coupled resonances are confined below 20 GHz, with optimal performance achieved beyond the reflecting waveguides' cutoff frequency. **b.i**, Over 180° of tuning is achieved across a broad frequency range between 21 and 30 GHz, covering the 5G New Radio band. **b.ii**, A close-in on the fine-tuning shows post-resonance reflectors achieve under 0.3° of resolution. **c,** Despite numerous switches loading the transmission lines, insertion loss through this device averages under 5 dB across the band. **d**, The return loss ($S_{11}$ and $S_{22}$) is over 20 dB, implying optimal impedance matching of this passive, reciprocal device.

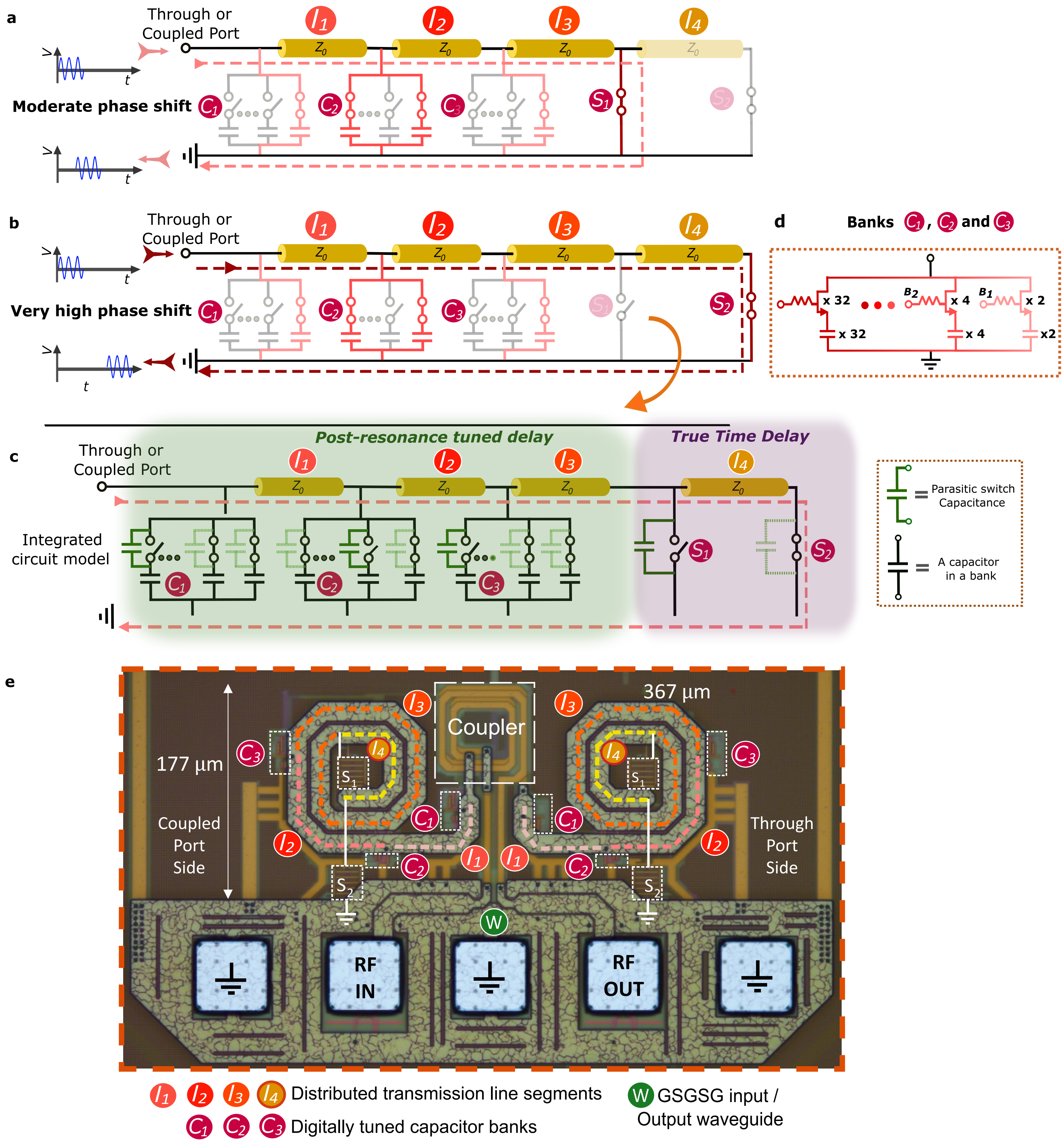

**Extended Data Figure 1. An extention in phase tuning range, with higher resolution, through a tunable-length return path in the reflectors of the Post-Resonance tuned phase shifter.** **a,** A moderate delay state is achieved by activating the basic load comprising waveguide segments $l_1$, $l_2$ and $l_3$ while disengaging segment $l_4$. **b,** A longer signal return path is activated when segment $l_4$ is brought into action. **c,** The on chip waveguide model consists of cascaded portions that combine Post-Resonance based delay and True-Time-Delay schemes. On-chip parasitic capacitances from the switches, in combination with the inductive segment $\boldsymbol{l_4}$, produce additonal coupled resonances and, thereby, additional phase shift. **d,** Capacitor banks consist of parallel legs of capacitors activated by N-type Metal Oxide Semiconductor transistor switches with progressively increasing sizes. **e,** A variant of the phase-shifter chip wherein the additional waveguide segment, $\boldsymbol{l_4}$, is coiled within the original footprint. Large microwave switches, $\boldsymbol{S_1}$ and $\boldsymbol{S_2}$, bring it in or out of action, providing an additional mechanism to reposition the coupled resonances of the non-uniform transmission lines.

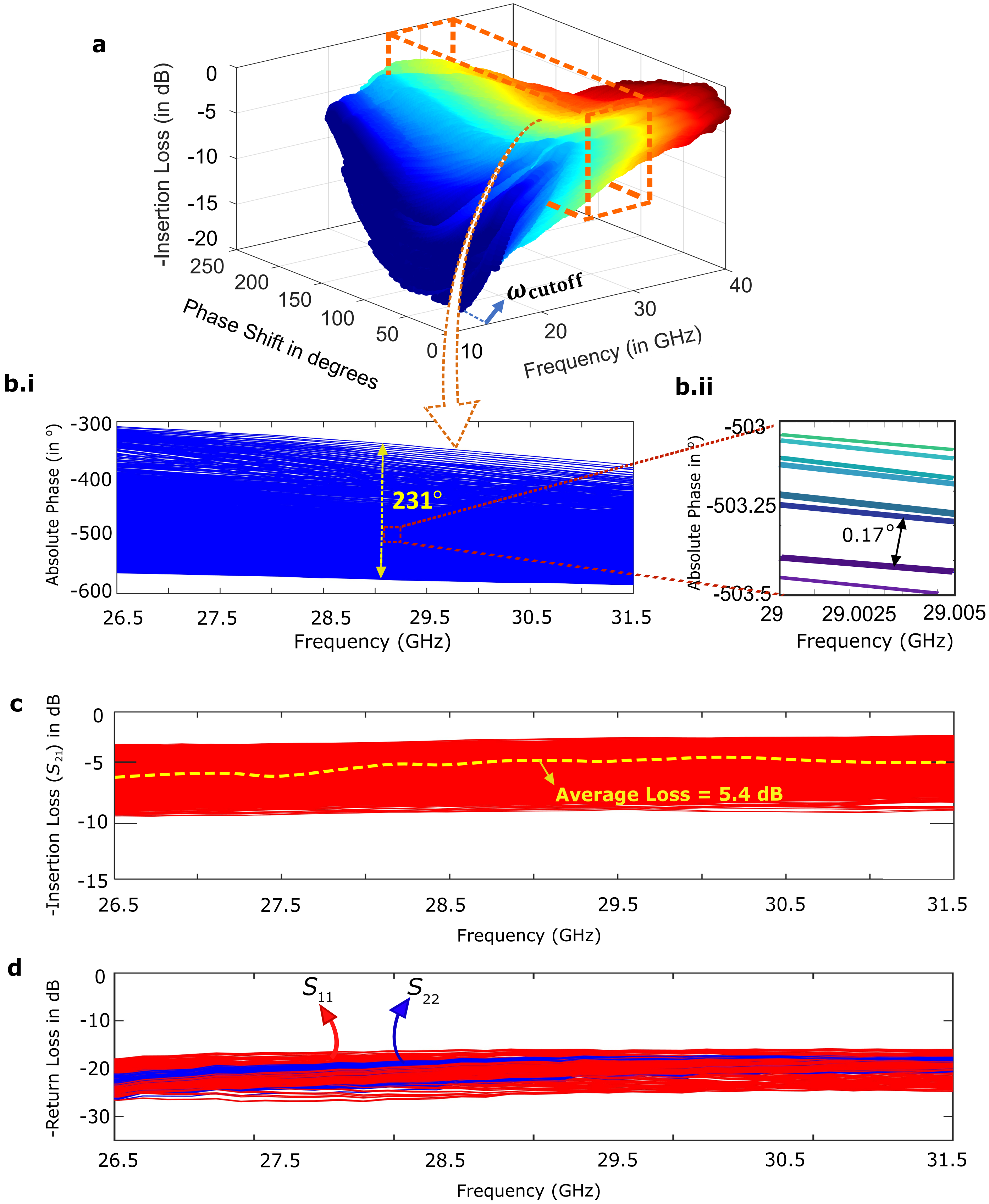

**Extended Data Figure 2. Measurements of the extended tuning range and finer resolution, using the same footprint as the original post-resonance tuned phase shifter. a,** The variation in phase shift and loss (vs frequency) measured for over 2000 combinations of 11-bit digital controls for the extended reflectors. As in the original phase shifter, loss decreases with increasing frequency, with mild tapering in the phase tuning range. **b.i,** Here, over 230° of phase tuning is achieved over the 26 - 32 GHz Multiple-Input Multiple-Output cell-phone band. **b.ii,** An ultra-fine resolution of under 0.17° is achieved. **c,** The insertion loss averages around 5.4 dB across the band despite using 34 switches in the circuit. **d,** The return loss to the input and output ports exceeds 20 dB, implying near-ideal impedance matching of the phase shifter.

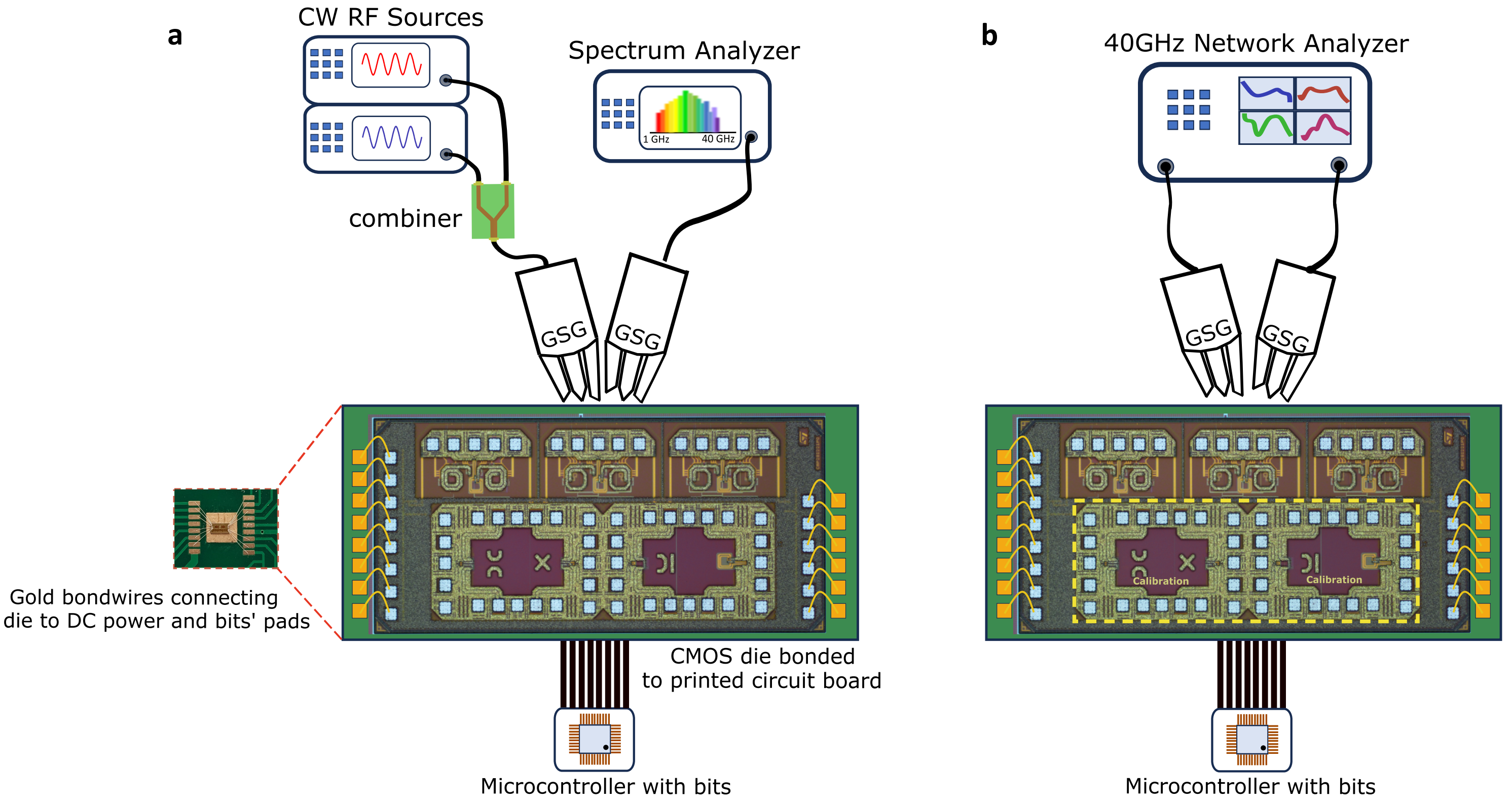

**Extended Data Figure 3. Test setup schematic for post-resonance tuned phase shifters.** The CMOS chip is bonded to a Printed Circuit Board. Gold wire-bonds connects its pads to logic supply voltages and digital input bits. Millimeter wave Ground-Signal-Ground-Signal-Ground probes transfer RF power into and out of the chip. **a,** RF input power is swept using power-combined synthesizers and the output spectrum is sensed on a spectrum analyzer. The third order intermodulation products are measured to gauge linearity of the device. **b,** Small-signal scattering parameters are measured with a network analyzer to characterize loss and phase-tuning states.

**Intermodulation products' characterization for the primary phase shifter (main article)**

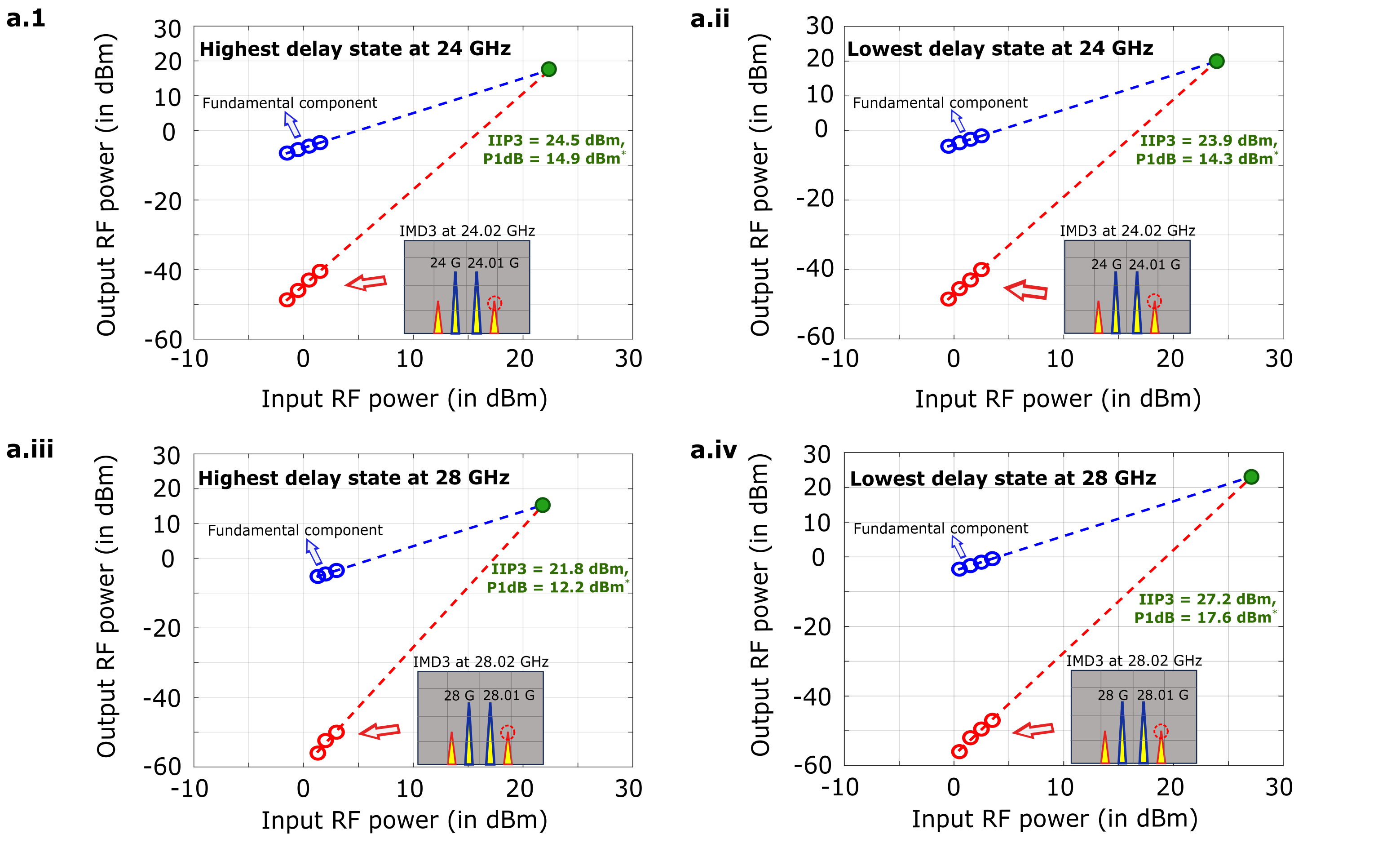

**Intermodulation products' characterization for the extended reflector variation (Extended Data Fig. 1)**

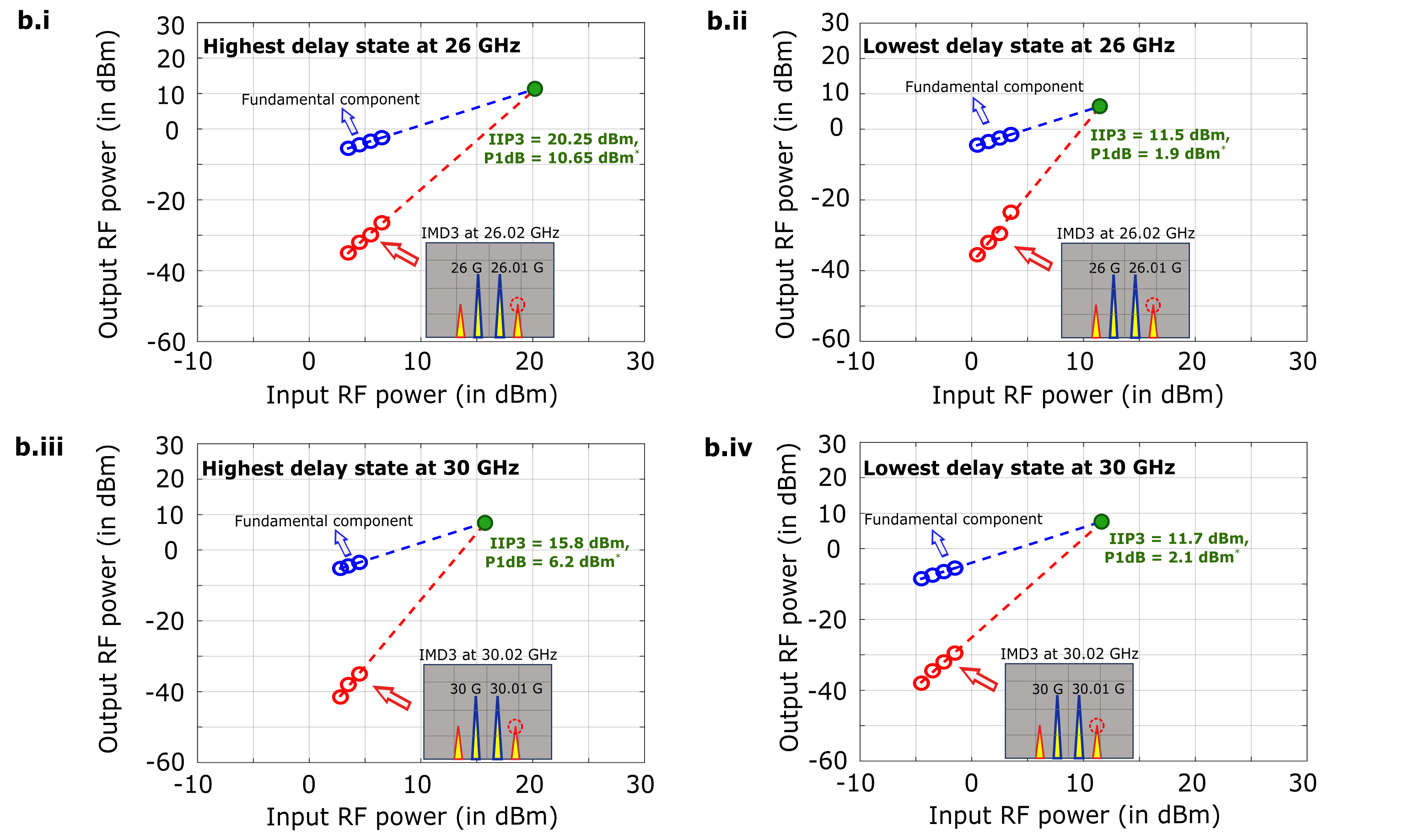

*** 1 dB compression point is calculated as IP1dB = IIP3 - 9.6 dB**

**Extended Data Figure 4. a,** Measured variations in linearity for the primary phase shifter (main article), across frequency and for high and low phase delay states at 24 GHz (**a.i** and **a.ii**) and at 28 GHz (**a.iii** and **a.iv**). Similar measurements for the second chip variant (with optional true-time-delay) at 26 GHz (**b.i** and **b.ii**) and at 30 GHz (**b.iii** and **b.iv**). Large switches that tune the length of the reflectors slightly boost third order intermodulation products.

# Supplementary Information Guide

## TABLE OF CONTENTS

## SUPPLEMENTARY NOTES' TITLES

**Supplementary Note 1.** Direct calculation of reflection coefficient for the three-segment load

**Supplementary Note 2.** Effects of finite quality factors for the multi-resonant load, using Bloch impedance-based filter design

**Supplementary Note 3.** Generalized coupled-mode model for scattering parameters

**Supplementary Note 4.** Visualizing effects of finite phase resolution and plotting 3D beam patterns

## SUPPLEMENTARY FIGURES' TITLES AND LEGENDS

**Fig S1.** Circuit to calculate reflection coefficients for the cascaded $\pi$ segments in the post-resonance load.

**Fig S2.** Direct analytical simulation of reflection coefficients' magnitude and phase for two quality factors (5 and 10) of the inductive segments in the reflector waveguides.

**Fig S3.** Representative filter characterization for **a,** a cascade of four LC unit cells using **b,** Bloch impedance, **c,** phase shift and **d** propagation constant through the periodically loaded transmission line used in the post-resonance reflector. All capacitances are identically swept, and each inductance equals 350 pH. Low Quality Factors of inductors ensure slow variations in input impedance and a large tuning range of phase shift (in this example, between 30 and 40 GHz).

**Fig S4.** Filter characterization for variations Bloch impedance, phase shift and propagation constant through the periodically loaded transmission line used in the post-resonance reflector, similar to Fig. S6, but with inductor Quality Factor = 10. While a higher Quality Factor ensures sharp frequency selectivity, it also introduces greater signal distortion due to high dispersion and sharp phase variations.

**Fig S5.** A coupled-mode model representing the multi-resonant waveguide used as a reflector. It is characterized by mode-coupling coefficients, internal dissipation, and coupling to an external port, where the reflection coefficient is computed. **b,** Broadband modulation of the first mode ($f_1$) influences the contributions of the resultant coupled resonances, affecting the magnitude of the reflection coefficient ($|S_{11}|$). This magnitude fluctuates rapidly at low frequencies and approaches near-ideal reflectivity at higher frequencies. **c,** Phase variations, for different values of $f_1$, indicate uniform post-resonance phase shifts that are accumulated from the coupled resonances. These trends align with the measured states of the phase-shifter chips (see Fig 4 and Extended Data Fig 2). In this illustration, $f_1$ is tuned from 18 to 55 GHz, and $f_2$ and $f_3$ are kept constant at 15 and 35 GHz respectively. Also, Q's of all resonators equal 5. Coupling terms are $\beta_{12} = 0.1e^{i\pi/4}$, $\beta_{23} = 0.2e^{i\pi/6}$ and $\gamma^{ext}$ = 60 GHz.

**Fig. S6.** A comparison of beam patterns projected from phase-shifting arrays. These profiles are computed by imposing a realistic silicon chip area of 3.1 $mm^2$ to contain the phase shifters. Beams are fired in a direction of $20^o$ in both azimuth and elevation. Post-resonance reflectors overcome the effects of insertion loss, reduced directivity and sharp sidelobes on the image.

**Fig. S7.** Variations of beam patterns from expanded arrays of post-resonance reflector-based delay elements. **a,** Arrays of 30 X 30 elements with much higher main lobe directivity than in Fig. S6 **b,** Ultra-sharp directivity for 10-bit arrays of 60 X 60 elements wherein sidelobes have nearly no influence.

## SUPPLEMENTARY TABLES' TITLES AND LEGENDS

**Supplementary Table 1|** An extended comparison table for state-of-the-art microwave and millimeter wave phase-shifters fabricated in CMOS, BiCMOS, GaN, GaAs and MEMS processes.

## SUPPLEMENTARY INFORMATION REFERENCES

**Supplementary Table 1|** A comparison table for state-of-the-art microwave and millimeter-wave phase-shifters in CMOS, BiCMOS, GaN, GaAs and MEMS technologies (active type in pink, passive type in green)

| Semiconductor Technology | Type of circuit | Chip area and number of elements per 1 mm² | Frequency Range | Average Insertion loss or gain | DC Power | Phase Tuning Range and Resolution | Signal Linearity | Reference and year |
|---|---|---|---|---|---|---|---|---|
| 130 nm CMOS | Active | 0.14 mm²<br>7 elements | 6 - 18 GHz | -2.1 dB<br>(with gain) | 8.7 mW | 338° range<br>22.5° resolution | Input P1dB =<br>-5.4 dBm | (*S2*)<br>2007 |
| 130 nm CMOS | Active | 0.14 mm²<br>7 elements | 15 – 26 GHz | -4.6 dB<br>(with gain) | 11.7 mW | 338° range<br>22.5° resolution | Input P1dB =<br>-0.8 dBm | (*S2*)<br>2007 |
| 130 nm BiCMOS | Active | 3.9 mm²<br>Larger than chip | 8-12 GHz | 14 dB<br>(with gain) | 195 mW | 180° range<br>6.25° resolution | Input P1dB =<br>-15 dBm | (*17*)<br>2017 |
| 180 nm CMOS | Active | 0.4 mm²<br>2 elements | 2-3 GHz | 1.5 dB<br>(with gain) | 24 mW | 360° range<br>33° resolution | Input P1dB =<br>-13.5 dBm | (*18*)<br>2010 |
| 65 nm CMOS | Active | 0.26 mm²<br>3 elements | 6-20 GHz | 13 dB<br>(with gain) | 137 mW | 360° range<br>2.8° resolution | Not measured | (*S3*)<br>2023 |
| 65 nm CMOS | Active | 0.4 mm²<br>2 elements | 51 - 64 GHz | 5 dB<br>(with gain) | 20.3 mW | 360° range<br>1.4° resolution | Not measured | (*S4*)<br>2019 |
| 180 nm BiCMOS | Active | 0.9 mm²<br>1 element | 6-18 GHz | 16.5 dB<br>(with gain) | 61 mW | 360° range<br>11.2° resolution | Not measured | (S5)<br>2010 |
| 130 nm BiCMOS | Active | 0.36 mm²<br>2 elements | 30-38 GHz | 1 dB<br>(with gain) | 5.4 mW | 360° range<br>22.5° resolution | Not measured | (*S6*)<br>2007 |
| 180 nm BiCMOS | Active | 0.19 mm²<br>5 elements | 15-35 GHz | -5 dB to +13 dB<br>(with gain) | 25 mW | 360° range<br>22.5° resolution | Input P1dB =<br>-6.25 dBm | (*19*)<br>2013 |
| 45 nm CMOS | Active | 0.19 mm²<br>16 elements | 40-65 GHz | -8 dB to -4 dB<br>(with gain) | 22.2 mW | 300° range<br>22.5° resolution | Input P1dB =<br>1.5 dBm | (S20)<br>2012 |
| 28 nm CMOS | Active | 0.01 mm²<br>100 elements | 106-122 GHz | 4 dB<br>(with gain) | 8.7 mW | 120° range<br>7° resolution | Not measured | (S21)<br>2017 |
| SiGe:C BiCMOS | Active | 0.03 mm²<br>33 elements | 86-106 GHz | 2.3 dB<br>(with gain) | 49 mW | 360° range<br>5.6° resolution | -3.4 dBm | (S22)<br>2023 |
| 90 nm CMOS | Passive | 0.08 mm²<br>12 elements | 50-65 GHz | 6.3 dB loss | 0 mW | 90° range<br>11.25° resolution | Input P1dB =<br>4 dBm | (S23)<br>2009 |
| 65 nm CMOS | Passive | 0.23 mm²<br>4 elements | 28-30 GHz | 10 dB loss | 0 mW | 360° range 30°<br>resolution | Not measured | (S24)<br>2020 |
| 28 nm<br>CMOS | Passive | 0.08 mm²<br>12 elements | 44-52 GHz | 7.5 dB loss | 0 mW | 90° range<br>1.5° resolution | Not measured | (*S7*)<br>2020 |
| 45 nm<br>SOI CMOS | Passive | 0.9 mm²<br>1 element | 27-31 GHz | 9.25 dB loss | 0 mW | 360° range<br>11.2° resolution | Not measured | (*S8*)<br>2019 |
| 65 nm<br>CMOS | Passive | 0.4 mm²<br>2 elements | 27-42 GHz | 12 dB loss | 0 mW | 360° range<br>11.2° resolution | Input P1dB =<br>-13.5 dBm | (*S9*)<br>2020 |
| 90 nm<br>CMOS | Passive | 0.3 mm²<br>3 elements | 36-40 GHz | 20.2 dB loss | 0 mW | 360° range<br>22.5° resolution | Not measured | (*S10*)<br>2017 |
| GaAs pHEMT | Passive | 0.65 mm²<br>1 element | 36-40 GHz | 10.7 dB loss | 0 mW | 360° range<br>22.5 ° resolution | Not measured | (*S11*)<br>2019 |
| ***28 nm<br>CMOS*** | ***Passive*** | ***0.08 mm²<br>12 elements*** | ***29-37 GHz*** | ***12.8 dB loss*** | ***0 mW*** | ***360° range<br>22.5 ° resolution*** | ***Not<br>measured*** | ***(20)<br>2020*** |
| 32 nm<br>SOI CMOS | Passive | 0.1 mm²<br>10 elements | 55-65 GHz | 5.5 dB loss | 0 mW | 360° range<br>45° resolution | Not measured | (*21*)<br>2013 |
| GeTe Phase-<br>Change Material | Passive | 0.43 mm²<br>2 elements | 26-34 GHz | 4-4.5 dB loss | 0 mW | 170° range<br>22.5° resolution | Not measured | (*S12*)<br>2020 |
| 90 nm<br>CMOS | Passive | 0.08 mm²<br>12 elements | 50-65 GHz | 6.25 dB loss | 0 mW | 90° range<br>22.5° resolution | Input P1dB =<br>4 dBm | (*22*)<br>2009 |
| GaAs, GaN<br>HEMTs | Passive | 2.5 mm² | 6-12 GHz | 7.4 dB loss | 0 mW | 167° range<br>50° resolution | Not measured | (*31*)<br>2022 |

**Supplementary Table 1 (continued)|** An extended comparison for state-of-the-art phase-shifters

| Semiconductor Technology | Type of circuit | Chip area and number elements in 1 mm² | Operating frequency | Average Insertion loss | DC Power | Phase Tuning Range and Resolution | Signal Linearity | Reference and year |
|---|---|---|---|---|---|---|---|---|
| ***250 nm BiCMOS*** | ***Passive*** | ***0.03 mm²<br>30 elements*** | ***24-30 GHz*** | ***4.5 dB loss*** | ***0 mW*** | ***180° range<br>45° resolution*** | ***Not measured*** | ***(23)<br>2020*** |
| 40 nm CMOS | Passive | 0.13 mm²<br>7 elements | 24-34 GHz | 6 dB loss | 0 mW | 315° range<br>45° resolution | Input P1dB =<br>11 dBm | (*24*)<br>2018 |
| ***65nm<br>CMOS*** | ***Passive*** | ***0.14 mm²<br>7 elements*** | ***32-40 GHz*** | ***17.5 dB loss*** | ***0 mW*** | ***360° range<br>3° resolution*** | ***Not measured*** | ***(34)<br>2020*** |
| 65 nm bulk CMOS | Passive | 0.16 mm²<br>6 elements | 24-34 GHz | 7.8 dB loss | 0 mW | 315° range<br>11.2° resolution | Not measured | (*34*)<br>2017 |
| Printed Circuit Board with varactors | Passive | Not applicable<br>Larger than a chip | 2.3-2.6 GHz | 3 dB loss | 0 mW | 360° range<br>36° resolution | Not measured | (*30*)<br>2018 |
| Printed Circuit Board with varactors | Passive | Not applicable<br>Larger than a chip | 3.5-3.9 GHz | 2.1 dB loss | 0 dB | 130° range<br>20° resolution | Not measured | (*33*)<br>2023 |
| MEMS (comb-drive actuators) | Passive | Not applicable<br>Larger than a chip | 500-550 GHz | 4 dB loss | 0 mW | 60° range (Avg.)<br>10° resolution | Not measured | (*38*)<br>2016 |
| MEMS switches on microstrip lines | Passive | 1.89 mm² | 70-86 GHz | 4.5 dB loss | 0 mW | 130° range<br>22.5° resolution | Not measured | (*S13*)<br>2022 |
| 65 nm CMOS | Passive | 0.24 mm² | 60-67 GHz | 14 dB loss | 0 mW | 360° range<br>2.8° resolution | Input P1dB =<br>10 dBm<br>(simulated) | (S29)<br>2023 |
| 65 nm CMOS | Passive | 0.033 mm² | 114-147 GHz | 17 dB loss | 0 mW | 360° range<br>5.6° resolution | Not measured | (S30)<br>2023 |
| 40 nm CMOS | Passive | 0.15 mm² | 26-32 GHz | 19 dB loss | 0 mW | 360° range<br>5.6° resolution | Input P1dB =<br>14 dBm | (S31)<br>2023 |
| 55 nm CMOS | Passive | 0.74 mm² | 14-18 GHz | 13 dB loss | 0 mW | 360° range<br>5.6° resolution | Not measured | (S32)<br>2024 |
| ***POST RESONANCE TUNING SCHEME*** | | | | | | | | |
| ***28 nm<br>SOI CMOS*** | ***Passive*** | ***0.064 mm²<br>16 elements*** | ***22-29 GHz*** | ***4.9 dB loss*** | ***0 mW*** | ***201° range<br>< 0.3° resolution*** | ***Input P1dB ><br>15 dBm*** | ***This design<br>(Main Article)*** |

**Rationale on selection of passive phase shifter designs representing the state-of-the-art.**

The marked designs used for comparison in the main article represent well-performing integrated circuit delay elements that are representative of different bit resolutions i.e. smallest phase shift. They also operate in the 20-40 GHz band of interest i.e., cell phone, satellite communication and MIMO frequency bands. Active phase shifters are discounted for comparison since they consume power and have poor linearity.

## Supplementary Note 1. Direct calculation of reflection coefficient for the three-segment load.

Each post-resonance tuned load is represented by cascaded CLC segments below in Fig S1. To model inductive segments more realistically, we assign internal series resistances $r_L$ for each (not shown in Fig S1). These impart a finite quality factor, $Q$, that equals $Q=\omega L/r_L$. On the CMOS chip, while each load is implemented as a coiled structure with all subsegments coupling to one another in some way, we will unwind the chain and represent it as linearly arranged LC segments for ease of analysis.

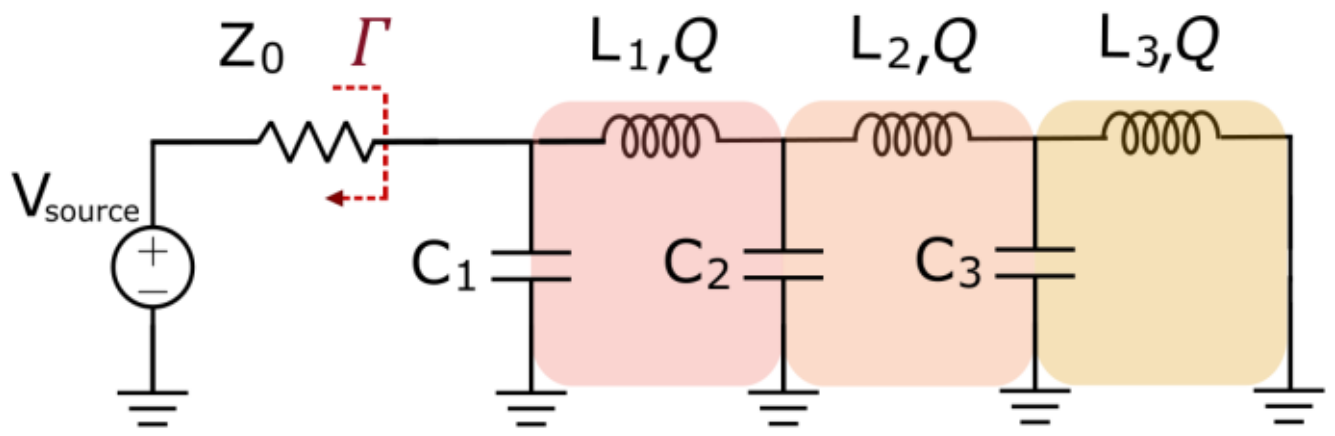

**Fig S1.** Circuit to calculate reflection coefficients for the cascaded $\pi$ segments in the post resonance load.

We recognize that the final segment is an $LC$ tank of impedance $Z_T = -\frac{j}{C_3\omega} || \left(j\omega L_3 + \frac{L_3\omega}{Q}\right)$. This would be in series with $\left(j\omega L_2 + \frac{L_2\omega}{Q}\right)$ and so forth, building a ladder. The exact expression for the reflection coefficient, using a characteristic impedance ($Z_0$ = 50 ohm), is then

$$\Gamma = \frac{Z_{in} - Z_0}{Z_0 + Z_{in}} = -\frac{Z_0 + Z_x}{Z_0 - Z_x} \qquad (S.1.1)$$

where

$$Z_x = \frac{j\left(j\omega L_1 + \frac{jL_1\omega}{Q} - \beta\right)}{C_1\omega\left(j\omega L_1 - \frac{j}{C_1\omega} + \frac{L_1\omega}{Q} - \beta\right)}, \qquad \text{where } \beta = \frac{j\left(j\omega L_2 + \frac{L_2\omega}{Q} - \varepsilon\right)}{C_2\omega\left(j\omega L_2 - \frac{j}{C_2\omega} + \frac{L_2\omega}{Q} - \varepsilon\right)} \qquad (S.2a)$$

and

$$\varepsilon = \frac{j\left(j\omega L_3 + \frac{L_3\omega}{Q}\right)}{C_3\omega\left(j\omega L_3 - \frac{j}{C_3\omega} + \frac{L_3\omega}{Q}\right)} \qquad (S.2b)$$

The magnitude and phase shift given by this formulation is shown for a few variations in capacitor banks and quality factors in Fig S.2. Reflection coefficient magnitudes in Figs S.2.a.i and S.2.b.i show that while low quality factor (Q=5) of passive structures encounters absorption losses at low, out of band frequencies, it facilitates an easier transition into a low-loss post-resonance regime. Also, when the quality factor is different (for example, Q=10), as Figs S.2.a.ii and S.2.b.ii show, the spread in the phase shift for various states of capacitor-banks is not maintained. In practice, it is beneficial to characterize several states and select ones that meet a target specification for both tolerable loss and wide tuning range, as shown in measurements in the main article.

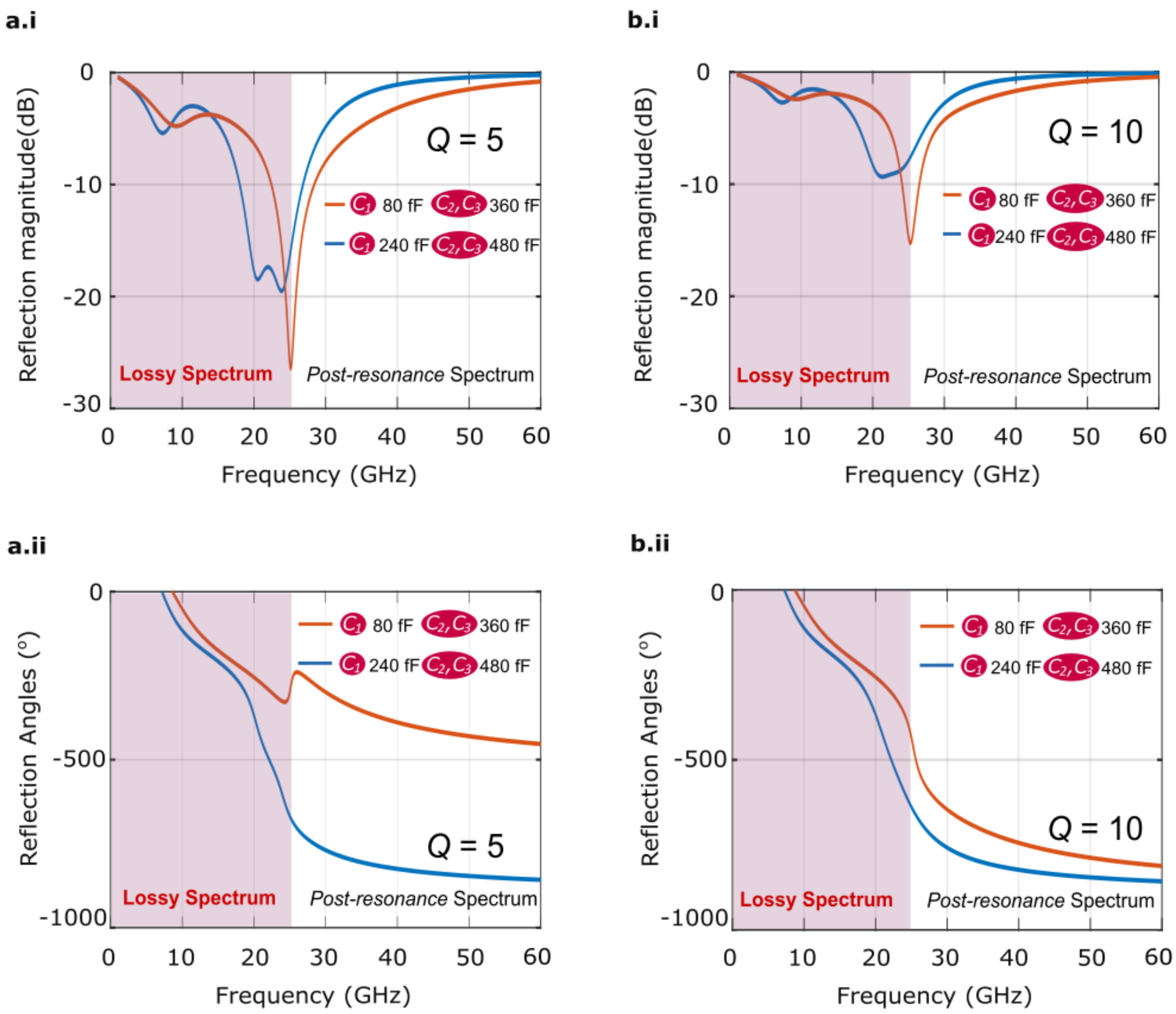

**Fig S2.** Direct analytical simulation of reflection coefficients' magnitude and phase for two quality factors (5 and 10) of the inductive segments in the reflector waveguides.

## Supplementary Note 2. Effects of finite quality factors for the multi-resonant load, using Bloch impedance-based filter design

The design of the load as a periodically loaded transmission line can also be performed by specifying its Bloch impedance [S26, S27], phase error across the band, and dispersion properties. Ideally, the load should suffer no rapid variations in input impedances and round-trip phase shifts across a wide band. Also, impedance of this filter-like structure must monotonically increase post-resonance to act as a good reflector. As the circuit in Fig S1 shows, the unit cells cascaded to form the full transmission line load in the device implemented in the main article are different. However, for simplicity, to study the effects of finite quality factors on chip, we will use the circuit in Fig S3.a which has identical inductive impedances, $Z_L = j\omega L$ with $Q = j\omega L / r_L$, and identical capacitive impedances, $Z_C = 1/j\omega C$. A transfer matrix may be used to analyze the behavior of each segment of the transmission line. The overall transfer matrix is obtained by cascading the transfer matrices of individual segments. Transfer matrices for the inductors and tunable capacitors are

$$T_L = \begin{pmatrix} 1 & Z_L + r_L \\ 0 & 1 \end{pmatrix} \qquad (S3a)$$

and

$$T_C = \begin{pmatrix} 1 & 0 \\ 1/Z_C & 1 \end{pmatrix} \qquad (S3b)$$

The overall transfer matrix for *N* cascaded sections and the final inductive section is

$$T_{\text{Load}} = (T_L \times T_C)^N \times T_L \qquad (S4)$$

The Bloch impedance is obtained from $T_{\text{Load}} = \begin{pmatrix} A & B \\ C & D \end{pmatrix}$ and is defined as the input impedance of an infinite periodic structure. For an infinite periodic structure, the input voltage $V_{\text{in}}$ and input current $I_{\text{in}}$ at the beginning of one period relate to the output voltage $V_{\text{out}}$ and current $I_{\text{out}}$ at the end of the period as

$$\begin{pmatrix} V_{\text{out}} \\ I_{\text{out}} \end{pmatrix} = \begin{pmatrix} A & B \\ C & D \end{pmatrix} \begin{pmatrix} V_{\text{in}} \\ I_{\text{in}} \end{pmatrix} \qquad (S5a)$$

where, for each *LC* segment, elements of the *T* matrix are

$$A = 1 + \frac{Z_L}{Z_0}, B = Z_L + Z_C, C = \frac{1}{Z_C + Z_0}, D = 1 + \frac{Z_C}{Z_0} \qquad (S5b)$$

Bloch impedance for this periodic structure comprising series *L*, shunt *C* unit cells can then be derived to be

$$Z_B = \frac{V_{in}}{I_{in}} = \frac{B}{1 - A} = \frac{1 - D}{C} = \sqrt{\frac{B}{C}} \qquad (S6)$$

Here, the propagation constant $\beta$ (or phase shift per unit cell) is related to the eigenvalues ($\lambda$) of the transfer matrix, $T_{\text{Load}}$ through the characteristic equation

$$\det(T - \lambda I) = 0 \qquad (S7)$$

This gives

$$\lambda_{1,2} = \frac{A \pm D \pm \sqrt{(A + D)^2 - 4(AD - BC)}}{2} = e^{\pm j\beta d} \qquad (S8)$$

The dispersion relation can be calculated by

$$\cos(\beta d) = \frac{A + D}{2} \qquad (S9)$$

For the capacitor-bank loaded reflective waveguide, Bloch impedance magnitude and phase shift show how the impedance of the transmission line varies with frequency, indicating resonant and anti-resonant frequencies. These help in identifying the operational frequency range for post resonance tuning and potential impedance mismatch. The dispersion relation shows how the phase constant varies with frequency. This provides insight into phase velocity and group velocity, factors critical to obtaining maximal-flatness filter response. These characteristics are illustrated for variations in quality factor and loading capacitors below (Fig S3 and Fig S4). In Fig S3.b, for a poor quality-factor of 3, we see that increasing capacitance (from 100 fF to 450 fF) introduces stronger resonant behavior, as seen by the peaks and troughs in the impedance and phase plots. Here, higher capacitance values lead to more pronounced reactive effects, increasing the overall impedance and causing more significant phase shifts (Fig S3.c). Higher capacitance also results in more variation in $\beta d$, indicating higher dispersive effects (Fig S3.d). The peaks and troughs in the plots for propagation constant indicate frequencies where the phase shift per cell is maximal. At these frequencies, the signal experiences significant phase change, which can lead to constructive or destructive interference depending on the overall length of the transmission line. A smooth plateauing of the propagation constant at higher frequencies indicates all these components will travel with similar phase velocity post resonance.

Now, in comparison, for a higher quality factor (Q = 10) in Fig. S4, we see more pronounced resonances with sharper impedance peaks and deeper troughs. These indicate stronger energy storage in the inductors and capacitors. The rapid changes in Bloch impedance across the band may cause issues in impedance matching, even when used in the post-resonance regime. The propagation constant also shows more variation, leading to more drastic phase shifts between coupled resonances and to drastically different delay through the waveguides' sub-segments. High dispersion implies that different frequency components will travel at different phase velocities, leading to signal distortion and potentially affecting the timing and integrity of signals, particularly in wideband applications spanning 10s of Gigahertz.

These observations indicate that, counterintuitively, for our application, a lower quality factor is preferred in the post-resonance regime for smooth, distinct phase variations and mild gradients in reflection coefficient magnitudes for different capacitor bank states.

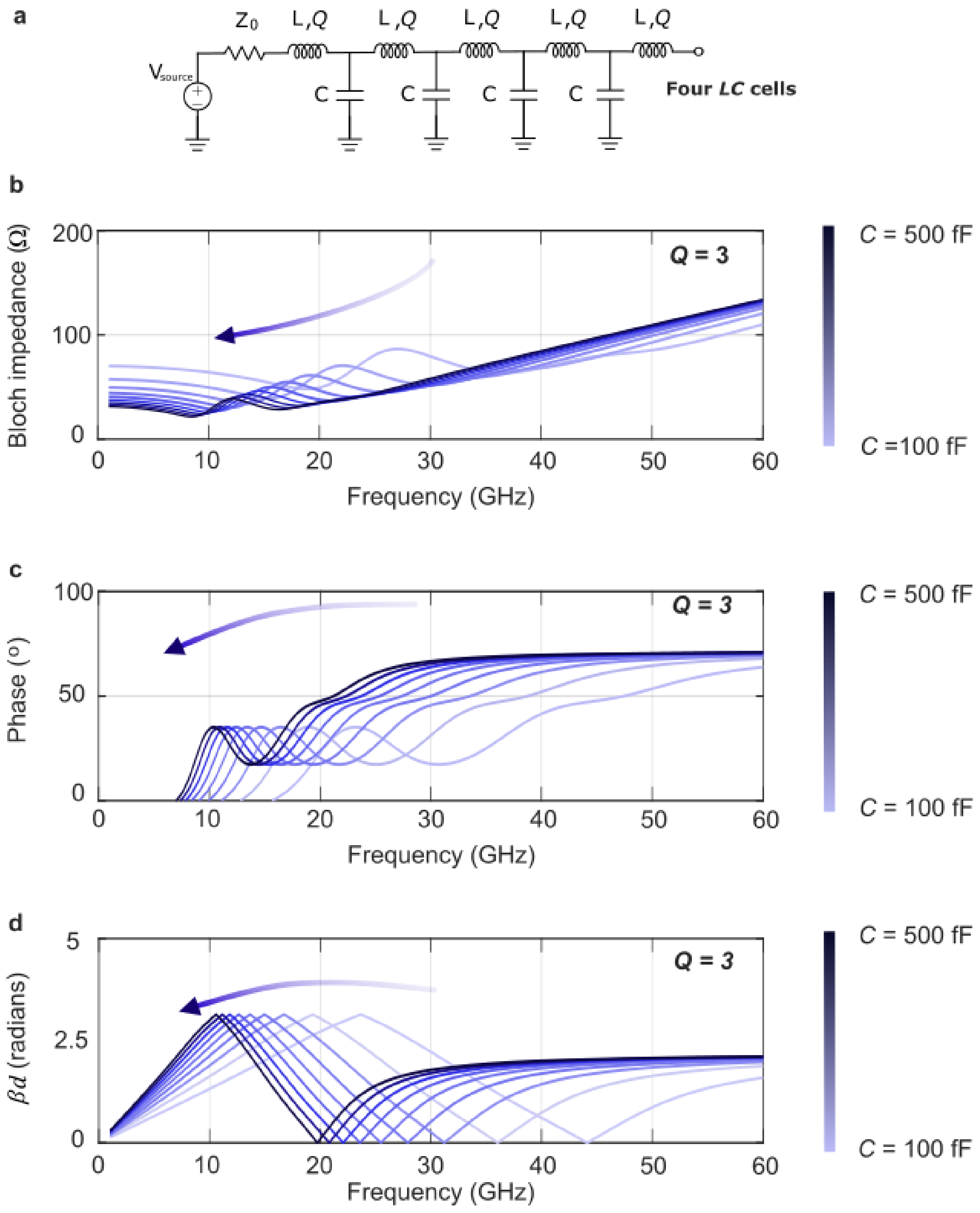

**Fig S3.** Representative filter characterization for **a,** a cascade of four LC unit cells using variations in **b,** Bloch impedance, **c,** phase shift and **d,** propagation constant through the periodically loaded transmission line used in the post-resonance reflector. All capacitances are identically swept, and each inductance equals 350 pH. Low Quality Factors of inductors ensure slow variations in input impedance and a large tuning range of phase shift (in this example, between 30 and 40 GHz).

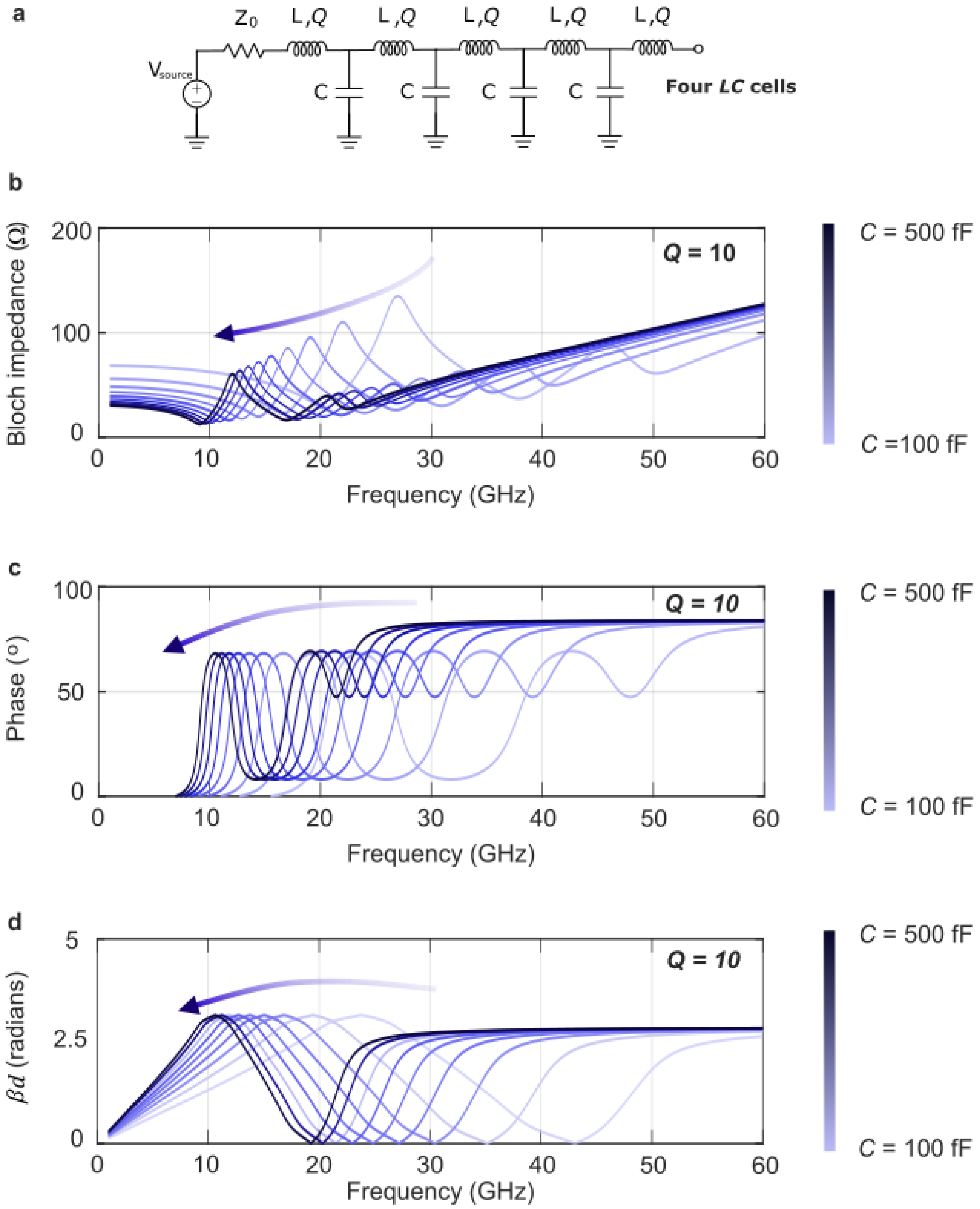

**Fig S4.** Filter characterization for variations in Bloch impedance, phase shift and propagation constant ($\beta d$) through the periodically loaded transmission line used in the post-resonance reflector, similar to Fig. S6, but with higher inductor Quality Factor of 10. While a higher Quality Factor ensures sharp frequency selectivity, it also introduces greater signal distortion due to high dispersion and sharp phase variations. This is not conducive to application in wideband passive phase shifters.

## Supplementary Note 3. Generalized coupled-mode model for scattering parameters

To quickly derive scattering parameters of the coupled resonators in the multi-resonant reflector, one may also consider it as a few coupled modes arranged linearly. It is driven at one end and is open at the other end. The frequency modes could be implemented in any form – as discrete inductor-capacitor $\pi$-shaped sections, traveling wave structures, coupled cavities, ring resonators, or otherwise. We will employ the coupled-mode theory framework for this formulation.

Consider a set of resonant modes with natural frequencies $\{\omega_1, \omega_2 \ldots \omega_n\}$ that have classical mode amplitudes $\{a_1, a_2 \ldots a_n\}$ . The time-evolution of the complex amplitudes $a_j(t)$ , where $j$ spans the set or resonators, is [S14, S15]

$$\frac{da_j}{dt} = -i\left(\omega_j - i\frac{\gamma_j}{2}\right)a_j(t) - i\sum_{k\neq j} c_{jk}(a_k + a_k^*) + \sqrt{\gamma_j^{\text{ext}}}a_j^{in} \qquad (S10)$$

Here, $a_j^{in}$ and $\gamma_j$ are the mode's external drive and dissipation terms respectively. $\gamma_j$, in turn, is a sum of the mode's internal loss, $\gamma_j^{int}$, and external loading, $\gamma_j^{ext}$, that provides a connection to signal ports. The dissipation rate of a resonator with a natural frequency $\omega_j$ can be expressed in terms of its quality factor as

$$Q_j = \frac{\omega_j}{\gamma_j} \qquad (S11)$$

It is more convenient to study the dynamics of this system in the frequency domain. Relabeling the Fourier variable $\omega$ as $\omega_j^s$ to signify the frequency of the input drive signal at which we'd like to study the response of the system, we write the internal mode amplitudes as a set $\vec{v} = \{a_j[\omega_j^s]\}$. The behavior of coupling terms between these modes is captured by the Equations of Motion matrix, **M,** and the environment-coupling matrix, **K**, that represents their connection to external ports.

$$K \equiv \text{diag}\left\{\sqrt{\gamma_j^{ext}}\right\} \qquad (S12)$$

We could then relate the coupled modes' response to the input drive as

$$-i\gamma_0\mathbf{M}\vec{v} = \mathbf{K}\vec{v}_{\text{in}} \qquad (S13)$$

where $\vec{v}_{\text{in}} = \{a_j^{in}[\omega_j^s]\}$ is the set of drive terms to each mode and $\gamma_0$ is an overall normalization rate given by

$$\gamma_0 \equiv \sqrt[N_P]{\prod_{k\in P}\gamma_k} \qquad (S14)$$

$N_P$ is the total number of external ports the ensemble of modes is connected to.

The matrix $\mathbf{M}$, when expanded, is

$$\mathbf{M} = \begin{bmatrix} \Delta_1 & \beta_{12} & \cdots & \beta_{1N} & \beta_{11^*} & \cdots & \beta_{1N^*} \\ \beta_{21} & \Delta_2 & \cdots & \beta_{2N} & \beta_{21^*} & \cdots & \beta_{2N^*} \\ \vdots & \cdots & \ddots & \vdots & \vdots & \ddots & \vdots \\ \beta_{N1} & \cdots & \cdots & \Delta_N & \beta_{N1^*} & \cdots & \beta_{NN^*} \\ \beta_{1^*1} & \cdots & \cdots & \beta_{1^*N} & -\Delta_1^* & \cdots & \beta_{1^*N^*} \\ \vdots & \cdots & \cdots & \vdots & \vdots & \ddots & \ddots \\ \beta_{N^*1} & \cdots & \cdots & \beta_{N^*N} & \beta_{N^*1^*} & \cdots & -\Delta_N^* \end{bmatrix} \tag{S15}$$

with diagonal detuning terms given by

$$\Delta_k \equiv \frac{1}{\gamma_0}\left(\omega_k^s - \omega_k + i\frac{\gamma_k}{2}\right) \tag{S16}$$

and off-diagonal coupling terms given by

$$\beta_{jk} \equiv \frac{c_{jk}}{2\gamma_0} \tag{S17}$$

where $c_{jk}$ are the coupling rates between modes $j$ and $k$. The scattering parameters are related to these matrices by [S16]

$$\mathbf{S} = i\frac{1}{\gamma_0}\mathbf{K}\mathbf{M}^{-1}\mathbf{K} - \mathbb{I} \tag{S18}$$

Since the three-resonator reflector in the delay element is passive and not parametrically driven, the Equations of Motion matrix of the reflector in Fig 3 of the main article with only three modes is represented by

$$\mathbf{M}_{\mathbf{reflector}} = \begin{bmatrix} \Delta_1 & \beta_{12} & \beta_{13} \\ \beta_{21} & \Delta_2 & \beta_{23} \\ \beta_{31} & \beta_{32} & \Delta_3 \end{bmatrix} \tag{S19}$$

Since this reflective waveguide is a reciprocal network, $\beta_{12}=\beta_{21}$ and $\beta_{23} = \beta_{32}$. Also, as the terminal resonators are not directly connected to one another, $\beta_{13}$ and $\beta_{31}$ are identically zero. Further, there is a high impedance mismatch, i.e., a short to ground, at one end of the chain of resonators and only the first resonator is given an input drive. Therefore, the scattering parameter of interest is the reflection coefficient, $S_{11}$. Its magnitude and phase responses are calculated using this framework and is shown in Fig S5. We see that by modulating the position of the resonance $\omega_1$, we affect the behavior of multiple coupled modes and can achieve different broadband phase states. This model predicts loss and phase characteristics that closely resemble the trends observed in experimental measurements (Fig 4) and circuit simulations (Fig 3c) wherein the rapid variations in reflection coefficient amplitude subside at lower frequencies, allowing reflectivity to recover at higher frequencies. This would be especially useful for pulse phase modulators for high-frequency qubit controllers [S17].

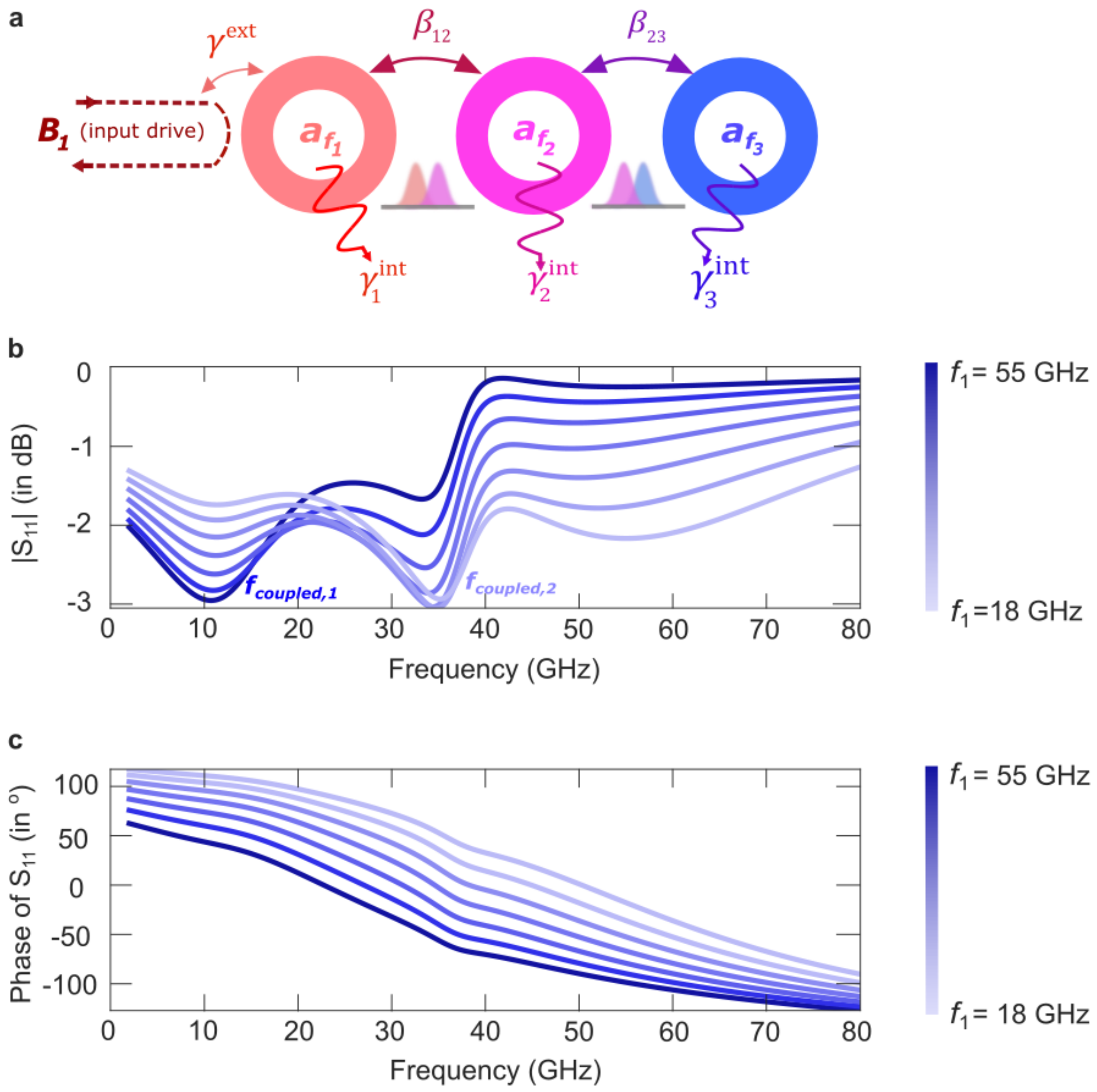

**Fig S5. a,** A coupled-mode model representing the multi-resonant waveguide used as a reflector. It is characterized by mode-coupling coefficients, internal dissipation, and coupling to an external port, where the reflection coefficient is computed. **b,** Broadband modulation of the first mode ($f_1$) influences the contributions of the resultant coupled resonances, affecting the magnitude of the reflection coefficient ($|S_{11}|$). This magnitude fluctuates rapidly at low frequencies and approaches near-ideal reflectivity at higher frequencies. **c,** Phase variations, for different values of $f_1$, indicate uniform post-resonance phase shifts that are accumulated from the coupled resonances. These trends align with the measured states of the phase-shifter chips (see Fig 4 and Extended Data Fig 2). In this illustration, $f_1$ is tuned from 18 to 55 GHz, and $f_2$ and $f_3$ are kept constant at 15 and 35 GHz respectively. Also, Q's of all resonators equal 5. Coupling terms are $\beta_{12} = 0.1e^{i\pi/4}$, $\beta_{23} = 0.2e^{i\pi/6}$ and $\gamma^{ext}$ = 60 GHz.

## Note 4. Visualizing effects of finite phase resolution and plotting 3D beam patterns

When a phased array is deployed in a practical communication scenario comprising multiple path lengths and reflections, a three-dimensional formulation of the array's beampattern is required. That would mean computation of the antennas' gain in the wavenumber-frequency space. The wave vector of the plane wave for the microwave radiation is given by

$$|\vec{k}| = \frac{\omega}{c} = \frac{2\pi}{\lambda} \qquad (S.20)$$

For a beam fired at an elevation angle, $\theta$, and at an azimuth angle, $\varphi$, in Cartesian coordinates,

$$k_x = k \sin\theta \cos\varphi$$
$$k_y = k \sin\theta \sin\varphi$$
$$k_z = k \cos\theta \qquad (S.21)$$

$k_x$, $k_y$ and $k_z$ are the rates of change of phase of the propagating wavefront in $\hat{x}$, $\hat{y}$ and $\hat{z}$ directions. Here we only consider isotropic antennas that radiate outward with a hemispherical profile. This is usually implemented with a back-baffle. The beampattern (BP) for all combinations of $k_x$, $k_y$ and $k_z$ is a combination of the spatial response of both the antennas, called the element pattern (EP), and the phase-shifters' Array Factor (AF). Using the expression for Array Factor from eqn. 3 (Methods),

$$BP = AF_{\text{phased array, actual}} \cdot EP \qquad (S.22)$$

To visualize this, we use the phased.URA routine in MATLAB to define the geometry of a phase-shifter array, with ideal models of antennas. We include effects of a finite number of tuning bits. The Array Factor, for a L×M array is first normalized and then Insertion Loss is uniformly subtracted across the full frequency band. Loss numbers are shown in Table S.2. A back-baffle is used to only permit radiation in one hemisphere. It is defined by the antenna object

```
antenna = phased.IsotropicAntennaElement (Frequency range, backbaffle…)
```

and requires installation of the Phased Array Toolbox. For simplification, quantized phase shifts in Fig 4b are rounded in the quantize_phase.m function used in Array_Factor_Code_Fig_4b.

Consider Fig. S6 that shows simulated 3D beam patterns. It accounts for the resolution, loss and size of each phase shifter when elements are arranged in arrays expanded to cover 3 mm$^2$ of on-chip area. Here, tradeoffs in design become evident. These arrays are fired at 20° in both azimuth and elevation. Passive vector modulated elements, with a high resolution of seven bits, can accommodate 5 × 5 arrays. They suffer heavy insertion loss and produce a feeble beam. A slight improvement in signal strength is made by switched filters which have a larger size but can only fit a 3 × 3 array and, therefore, have weak directivity that blurs the transmitted image. All-pass networks, it would seem, could compensate for both passive losses on chip and directivity problems because of their small size on chip (the chip could fit a 10×10 array). However, they produce numerous, strong sidelobes that could induce a high false alarm rate in radar, with the worst sidelobe that is only 4.5 dB weaker than the main lobe. This would need complex digital signal processing for error correction.

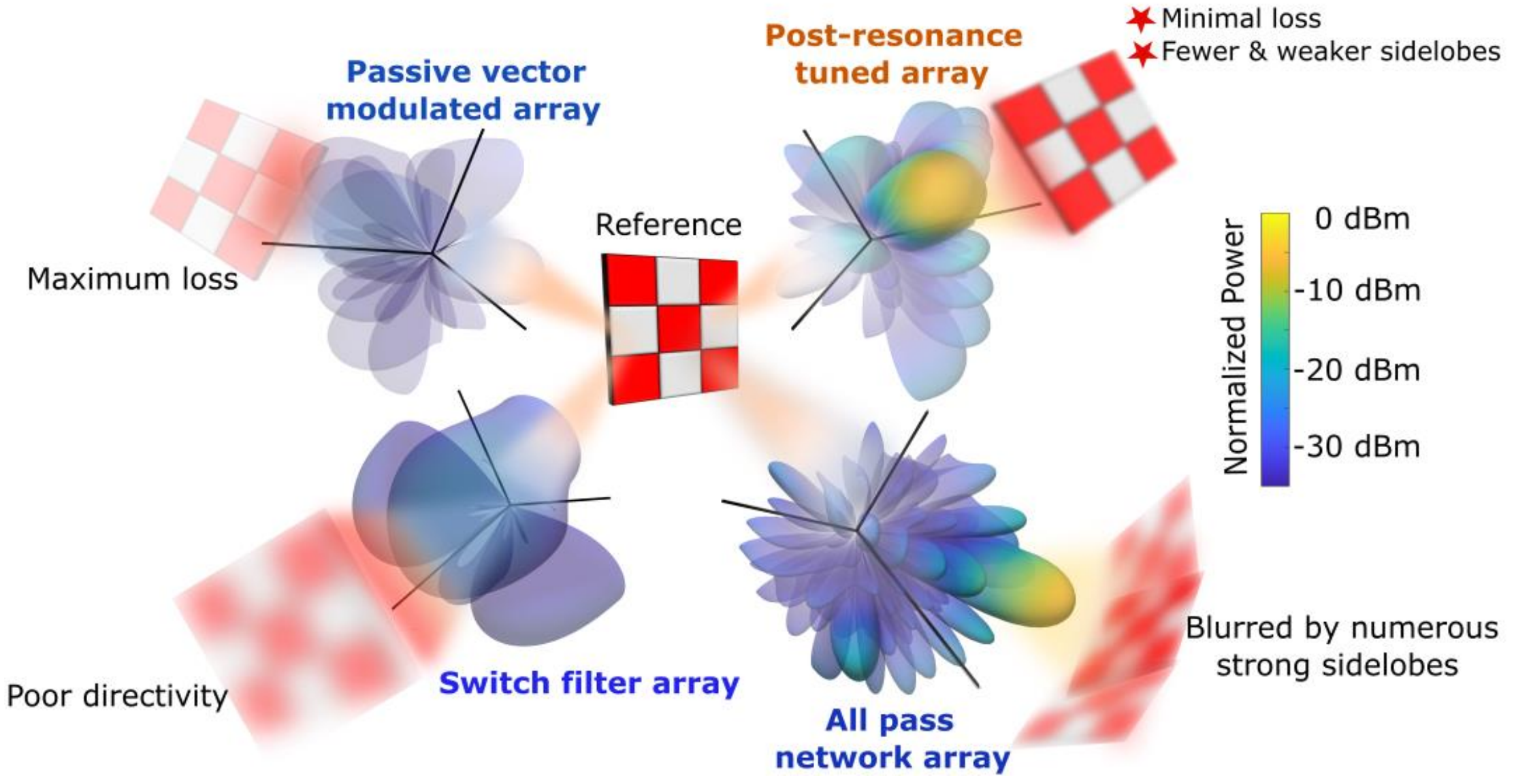

**Fig. S6.** A comparison of beam patterns projected from phase-shifting arrays. These profiles are computed by imposing a realistic silicon chip area of 3.1 mm$^2$ to contain the phase shifters. Beams are fired in a direction of 20° in both azimuth and elevation. Post-resonance reflectors overcome the effects of insertion loss, reduced directivity and sharp sidelobes on the image.

We find that the post-resonance tuned array has minimal signal loss, high resolution with far fewer sidelobes, with the worst sidelobe that is 13 dB weaker than the main lobe. This is consistent with observed array factors of high-resolution phase shifters (greater than four bits) that plateau at a suppression level of 13 dB [S19], but which have previously had high loss (> 8 dB). Its elements' miniature footprint can fit as many as 50 phase shifters on a 3 mm$^2$ chip. Of course, if these phase shifters are used in conjunction with amplitude weighting, sidelobe suppression can be further suppressed [S28]. To gauge how the beam pattern evolves by scaling the number of elements in an array of post-resonance tuned phase shifters is increased to 900 and 3600. As Fig. S7 shows, 60 × 60 arrays would give razor-sharp precision for wireless communication [S18]. MATLAB code for these patterns is found in the Zenodo repository.

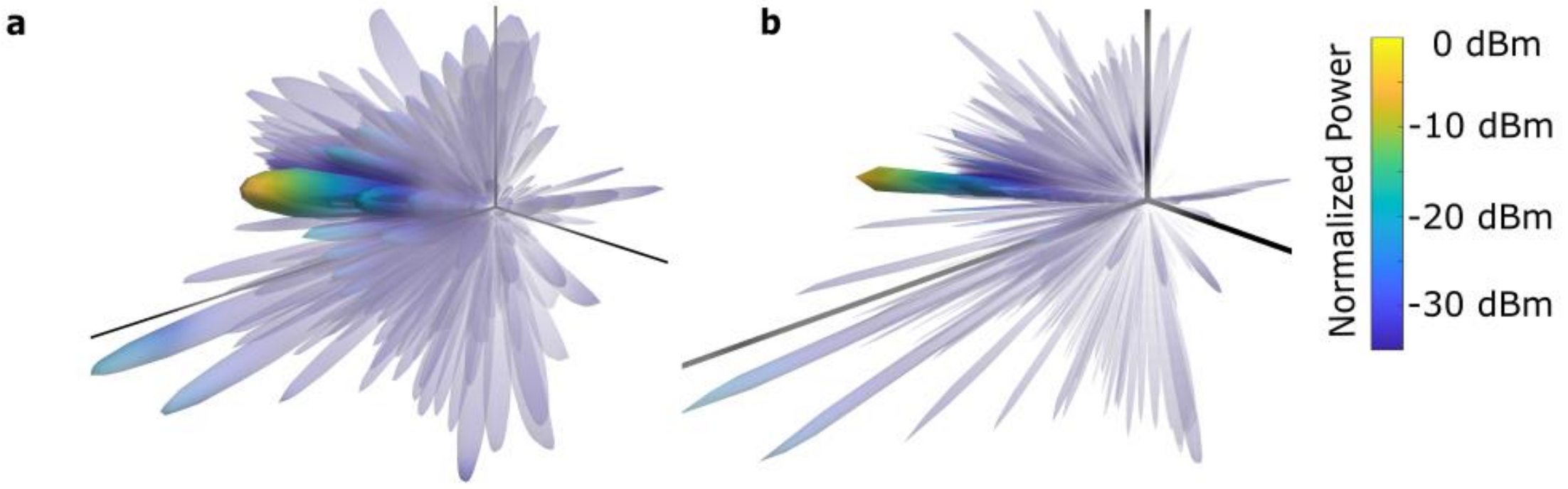

**Fig. S7.** Variations of beam patterns from expanded arrays of post-resonance reflector-based delay elements. **a,** Arrays of 30 X 30 elements with much higher main lobe directivity than in Fig. S6 **b,** Ultra-sharp directivity for 10-bit arrays of 60 X 60 elements wherein sidelobes have nearly no influence.